\documentclass[preprint,12pt,sort&compress]{elsarticle}
\usepackage[utf8]{inputenc}
\usepackage{color}
\usepackage{graphicx}
\usepackage{placeins}
\usepackage[margin=1in]{geometry}
\usepackage{appendix}
\usepackage{amsmath, amssymb, bm}
\usepackage{multirow}
\usepackage{placeins}
\usepackage{xfrac}
\usepackage[colorlinks=true, linkcolor=blue, citecolor=blue]{hyperref}
\usepackage[nameinlink, capitalize]{cleveref}
\usepackage[font=footnotesize, labelfont=bf]{caption}
\usepackage{subcaption}
\usepackage[version=4]{mhchem}
\usepackage{lineno}
\usepackage{bm}
\usepackage{comment}
\usepackage[T1]{fontenc}
\usepackage{anyfontsize}

\usepackage{float}

\journal{Journal of Nuclear Materials}

\begin{document}

\begin{frontmatter}

\title{Molecular-dynamics study of diffusional creep in uranium mononitride}

\author[ncsu]{Mohamed AbdulHameed}
\author[ncsu,inl]{Benjamin Beeler\corref{qwe}}
\cortext[qwe]{Corresponding author}
\ead{bwbeeler@ncsu.edu}
\author[lanl]{Conor O.T. Galvin}
\author[lanl]{Michael W.D. Cooper}
\author[ncsu]{Nermeen Elamrawy}
\author[west]{Antoine Claisse}

\address[ncsu]{Department of Nuclear Engineering, North Carolina State University, Raleigh, NC 27695}
\address[inl]{Idaho National Laboratory, Idaho Falls, ID 83415}
\address[west]{Westinghouse Electric Sweden, Västerås, SE 72163, Sweden}
\address[lanl]{Los Alamos National Laboratory, Los Alamos, NM 87545}

\begin{abstract}

Uranium mononitride (UN) is a promising advanced nuclear fuel due to its high thermal conductivity and high fissile density. Yet, many aspects of its mechanical behavior and microstructural features are currently unknown. In this paper, molecular dynamics (MD) simulations are used to study UN's diffusional creep. Nanometer-sized polycrystals are used to simulate diffusional creep and to calculate an effective GB width. It is found that Nabarro-Herring creep is not dominant in the temperature range of 1700--2000 K and that the dominant diffusional creep mechanism is Coble creep with an activation energy of 2.28 $\pm$ 0.09 eV. A method is proposed to calculate the diffusional GB width and its temperature dependence in polycrystals. The effective GB width of UN is calculated as 2.69 $\pm$ 0.08 nm. This value fits very well with the prefactor of the phenomenological Coble creep formula. It is demonstrated that the most comprehensive thermal creep model for UN can be represented as the combination of our Coble creep model and the dislocation creep model proposed by Hayes \textit{et al.}

\end{abstract}


\end{frontmatter}


\newpage

\section{Introduction}

Uranium mononitride (UN) is a potential next-generation nuclear fuel with several favorable properties like high fissile density, good heat transfer, compatibility with most potential cladding materials, and longer fuel residence periods \cite{Ekberg2018, Wallenius2020, Uno2020}. However, it also has challenges like complex manufacturing processes, high cost due to $^{15}$N enrichment, and vulnerability to steam at high temperatures \cite{Ekberg2018, Wallenius2020, Uno2020}. Many aspects of the mechanical behavior and microstructural features of UN at high temperatures and under irradiation are lacking in both consistent experimental data and detailed mechanistic understanding. The mechanical properties of nuclear fuels are crucial for a better understanding of the pellet-cladding mechanical interactions (PCMI) that occur during reactor operation, where cladding creep-down and fuel swelling lead to contact and the development of stresses between the cladding and the fuel, which, with the build-up of fission products, promotes stress corrosion cracking in the cladding material \cite{Frazer2021}. Fuel creep has an important role in accommodating void swelling once the fuel contacts the cladding \cite{Wallenius2020}.


There are a limited number of experimental investigations into the creep behavior of UN. Hayes \textit{et al.} \cite{Hayes1990II} developed an empirical correlation for the steady-state thermal creep rate of UN based on three experimental investigations by Fassler \textit{et al.} \cite{Fassler1965}, Vandervoort \textit{et al.} \cite{Vandervoort1968}, and Uchida and Ichikawa \cite{Uchida1973}. These experiments were conducted at a temperature range of 1373--2083 K, a stress range of 13--55.1 MPa, and for five grain sizes: 9, 15, 30, 140, and 2000 $\mu$m. In addition, the tested samples contained varying levels of carbon and oxygen impurities. To determine the mechanism controlling the creep of UN, Hayes \textit{et al.} \cite{Hayes1990II} determined the average stress exponent, based only on the data by Fassler \textit{et al.} \cite{Fassler1965} and Vandervoort \textit{et al.} \cite{Vandervoort1968}, to be $n = 4.5$. They therefore concluded that the dominant creep mechanism is most probably a climb-controlled dislocation glide mechanism, for which there is no grain size dependence, i.e., the grain size exponent $q = 0$. The data by Uchida and Ichikawa \cite{Uchida1973} gave a stress exponent $n$ = 1.8--2.6, which is indicative of a combination between diffusional and dislocation creep. However, Hayes \textit{et al.} decided to disregard this data set because its temperature range was intermediate between that of Fassler \textit{et al.} \cite{Fassler1965} and Vandervoort \textit{et al.} \cite{Vandervoort1968}, and they deemed it was unlikely that a diffusion-based mechanism would only be active in such a narrow temperature range (i.e., 1583--1773 K). The rest of the correlation (i.e., the prefactor and the activation energy) was only fit to the data by Vandervoort \textit{et al.} \cite{Vandervoort1968} because it is the only data set that reported stoichiometric samples of 100\% theoretical density. The final correlation reads:
\begin{equation}
\Dot{\epsilon} = 2.054 \times 10^{-3} \sigma^{4.5} \, \mathrm{exp} \! \left( - \frac{39369.5}{T} \right),
\label{Eq:Hayes}
\end{equation}
where $\Dot{\epsilon}$ is the creep rate measured in s$^{-1}$, $\sigma$ is the stress measured in MPa, and $T$ is the temperature measured in K. The correlation is strictly valid for $T$ = 1770--2083 K (although it was assumed by Hayes \textit{et al.} to yield reasonable estimates of the creep rate over the range of 298--2523 K), $\sigma$ = 20--34 MPa, and theoretically dense UN. The exponential term in \cref{Eq:Hayes} corresponds to an activation energy $Q$ = 3.39 eV.

Rogozkin \textit{et al.} \cite{Rogozkin2003} performed experiments and reported thermal creep data for U$_{0.8}$Pu$_{0.2}$N. They found that for theoretically dense samples, the thermal creep rate follows the correlation:
\begin{equation}
\Dot{\epsilon} = 27.7 \sigma^{1.35} \, \mathrm{exp} \! \left( - \frac{40000}{T} \right),
\label{Eq:UPuN}
\end{equation}
which indicates an activation energy $Q$ = 3.45 eV. The investigators noted that the mass fraction of oxygen and carbon impurities did not exceed 0.15\% and the stress varied from 10 to 60 MPa. However, they did not report any grain sizes or an explicit unit for the creep rate. Note that a stress exponent of 1.35 in \cref{Eq:UPuN} is close to 1 and might indicate a diffusion-dominant creep mechanism \cite{Rogozkin2003, Meyers2009}.

Nabarro-Herring (N-H) creep of UN has been investigated via computational tools. Kotomin \textit{et al.} \cite{Kotomin2009} conducted a DFT investigation of the diffusion mechanisms of U and N atoms in UN and calculated their associated migration energies. Assuming thermal creep is rate-limited by the diffusion of the slower species (i.e., uranium), they concluded that the activation energy of thermal creep in bulk UN is 5.6 eV. Their thermal creep model is incorporated into the TRANSURANUS fuel performance code \cite{Kotomin2009}.

Finally, Konovalov \textit{et al.} \cite{Konovalov2016} conducted a compilation of available data on UN, (U, Pu)N, UC, and mononitrides of refractory metals (e.g., Zr and Ti) to find a correlation for thermal and irradiation creep in UN and (U, Pu)N. Based on the data, they assumed a combination of dislocation-climb creep and Coble creep according to:
\begin{equation}
\Dot{\epsilon} = a_1 \sigma^n \, \mathrm{exp} \! \left( - \frac{Q_b}{k_\mathrm{B} T} \right) + f a_2 \frac{\sigma}{k_\mathrm{B} T} \, \mathrm{exp} \! \left( - \frac{ Q_\mathrm{GB}}{k_\mathrm{B} T} \right),
\end{equation}
where $a_1$ and $a_2$ are estimated from typical values for refractory-metal mononitrides, the stress exponent of the dislocation-climb creep part is $n=4.5$, the same as in \cref{Eq:Hayes}, and $f$ is the volume fraction of GBs. Note that the dependence of the Coble creep on the grain size is included in $a_2$. They further assumed that the activation energy for the dislocation-climb creep is equal to the activation energy of bulk diffusion of U atoms in UN, i.e., $Q_b$ = 5 eV. Based on the observation that in typical metallic compounds, the activation energy of GB diffusion is nearly half that of bulk diffusion, they assumed $Q_\mathrm{GB} = 0.53 Q_b$, similar to UC, because measurements of U self-diffusion at GBs in UN are not yet available. An important result of their analysis is that diffusional creep is the dominant mechanism in the temperature range of fuel manufacturing processes and nuclear reactor operation, although they did not clarify the effect of stress and grain size.

A few observations can be made about the previous studies. The assumption by Hayes \textit{et al.} that dislocation creep is the dominant creep process cannot be made for all grain sizes at all temperatures based only on very few experimental data points. The studies by Fassler \textit{et al.} \cite{Fassler1965} and Vandervoort \textit{et al.} \cite{Vandervoort1968}, on which the stress exponent in \cref{Eq:Hayes} was based, report grain sizes of 30, 140, and 2000 $\mu$m. However, typical grain sizes of manufactured UN fuel range between 5 and 30 $\mu$m \cite{Adachi2009, Johnson2018, Yang2021b, He2021}. Johnson and Lopes \cite{Johnson2018} report that the grain size for UN to achieve optimum fuel porosity (i.e., 0\% open porosity to suppress oxidation and 4\% closed porosity to accommodate gas fission products) is around 8 $\pm$ 1 $\mu$m. Uchida and Ichikawa \cite{Uchida1973}, who found a stress exponent closer to 1, reported grain sizes of 9 and 15 $\mu$m. That is, diffusional creep might be the dominant creep mechanism at the grain sizes typical of the manufactured fuel. The existence of diffusional creep as a contributing process is supported by the studies of Rogozkin \textit{et al.} \cite{Rogozkin2003} and Konovalov \textit{et al.} \cite{Konovalov2016}, and can be dominant if the activation energy of a diffusional creep mechanism is found to be lower than that of dislocation creep. Basing the thermal creep model of UN on bulk diffusion activation energies, as was done by Kotomin \textit{et al.} \cite{Kotomin2009}, is questionable, especially because GB diffusion is known to be much faster than bulk diffusion \cite{Mauro2020}, and the typical grain sizes of manufactured UN (5--30 $\mu$m) are quite small and might render Coble creep as the dominant mechanism at the temperature range of practical interest. Finally, Nabarro-Herring creep is expected to be dominant near the melting point \cite{Courtney2005}.

It is worth mentioning that Hayes \textit{et al.} \cite{Hayes1990III} also developed correlations for diffusion in UN based on a few scattered data points. Their estimated activation energies for U and N atoms are 0.69 and 1.66 eV, respectively, which raises suspicions because U atoms are not expected to have a lower diffusion activation energy than that of N atoms. Hayes \textit{et al.} reported that their correlation for N diffusion is for tracer diffusion measurements, and is only representative of extremely hyper-stoichiometric conditions. Finally, while the analysis by Konovalov \textit{et al.} \cite{Konovalov2016} is qualitatively appealing, it lacks quantitative insights because it is based on typical values and educated guesses. In summary, the literature on creep in UN is sparse, somewhat contradictory, and requires additional investigation to identify the dominant creep mechanisms at various temperatures, stresses, and grain sizes.


Molecular-dynamics (MD) studies have been used with success to study diffusional creep in, e.g., Si \cite{Keblinski1998}, Pd \cite{Yamakov2002}, and UO$_2$ \cite{Desai2008, Galvin2025}. These studies set up a simulation for idealized microstructures, where the supercells are equilibrated at a certain temperature and put under stress, and the strain rate is then analyzed. Haslam \textit{et al.} \cite{Haslam2004} further developed this approach by using random grain sizes and shapes, instead of idealized microstructures, to observe both creep rate and grain growth mechanisms in Pd. Cooper \textit{et al.} \cite{Cooper2021} used another approach (hereafter termed the parameter-based approach) for predicting Coble creep rates for U$_3$Si$_2$ where, instead of directly observing creep, an initial creep formula is assumed, and then, a lower-length scale modeling approach is used to calculate the formula parameters, e.g., defect volumes and GB diffusivities. To the best of our knowledge, no computational studies exist in the literature for Coble creep in UN.

In order to conduct MD simulations, a sufficiently accurate interatomic potential is required. Two promising interatomic potentials of UN exist in the literature: Tseplyaev and Starikov's angular-dependent potential \cite{Tseplyaev2016}, and Kocevski \textit{et al.}'s embedded-atom method (EAM) potential \cite{Kocevski2022II}. Both potentials have been utilized and compared in our previous works \cite{AbdulHameed2024, AbdulHameed2024b}, however, their ability to model dynamical processes like diffusional creep has not been assessed. In this work, nano-sized polycrystals are employed to perform MD simulations of diffusional creep in UN. The polycrystals are also used to study GB diffusion and the temperature dependence of the diffusional GB width in UN. A parameter-based model of Coble creep is also constructed and compared to the simulation model. 

\section{Methods}

All MD calculations performed in this work utilize the Large-scale Atomic/Molecular Massively Parallel Simulator (LAMMPS) software package \cite{Thompson2022} using a 1 fs time step and the Tseplyaev \cite{Tseplyaev2016} and Kocevski \cite{Kocevski2022II} force-field potentials. Periodic boundary conditions (PBCs) are applied to all supercells. The Atomsk code \cite{Hirel2015} is used to generate polycrystals by the Voronoi tessellation method. The OVITO software package \cite{Stukowski2010} is used for supercell visualization and analysis. 

\subsection{Creep mechanisms}

Steady-state creep strain rates can be represented by the Mukherjee-Bird-Dorn equation \cite{Mukherjee2002}: 
\begin{equation}
\Dot{\epsilon} = \frac{A}{T} \frac{\sigma^n}{d^q} \, \mathrm{exp} \! \left( - \frac{Q}{k_\mathrm{B} T} \right) \approx C \frac{\sigma^n}{d^q} \, \mathrm{exp} \! \left( - \frac{Q}{k_\mathrm{B} T} \right),
\label{Eq:MBDEq}
\end{equation}
where $T$ is temperature, $\sigma$ is stress, $d$ is the grain size, $Q$ is an activation energy, $k_\mathrm{B}$ is the Boltzmann constant, and $A$, $n$, and $q$ are dimensionless constants. The inverse temperature dependence $A/T$ is relatively weak, especially at temperatures where creep is important, and is usually neglected \cite{Dowling2020}. That is why $A/T \approx C$ is treated as a constant prefactor. The values of the exponents $n$ and $q$ depend on the creep mechanism under consideration. For example, for power-law (dislocation) creep, $n$ = 3--8 and $q$ = 0, i.e., no grain size dependence, whereas for diffusional creep, $n$ = 1 and $q$ = 2 (for the Nabarro-Herring mechanism) or $q$ = 3 (for the Coble mechanism).

It should be noted that the Harper-Dorn creep, a dislocation creep mechanism, has a stress exponent $n=1$, similar to diffusional-creep mechanisms, as well as $q=0$, i.e., no grain size dependence. However, for Harper-Dorn creep to be significant, the material must have a large grain size (in the order of 100 $\mu$m \cite{Meyers2009}); otherwise, diffusional creep mechanisms are more likely to dominate. There is little evidence for the existence of Harper-Dorn creep in ceramics \cite{Meyers2009}. Ceramics generally have smaller grain sizes, fewer available slip systems, and high Peierls-Nabarro stresses, all of which lead to the prevalence of other creep mechanisms \cite{Meyers2009}. For this reason, the Harper-Dorn creep is excluded from consideration as a competing creep mechanism in UN, and dislocation creep refers exclusively to power-law creep.

In this work, we study diffusional creep, whose rate, in general, is a sum of the Nabarro-Herring creep rate, $\Dot{\epsilon}_{\mathrm{NH}}$, and the Coble creep rate, $\Dot{\epsilon}_C$, since both mechanisms operate in parallel \cite{Courtney2005}.


For Coble creep, the steady-state strain rate is expressed as \cite{Coble1963, Courtney2005}:
\begin{equation}
\Dot{\epsilon}_C = A_C \frac{\sigma \Omega}{k_\mathrm{B} T} \frac{ D_{\mathrm{GB}} \delta }{d^3},
\label{Eq:Coble}
\end{equation}
where $A_C$ depends on the grain shape ($A_C = 46.347$ for spherical grains), $D_{\mathrm{GB}}$ is the \textit{effective} GB diffusivity, $\delta$ is the GB width, and $\Omega$ is the vacancy volume usually approximated as the atomic volume or expressed as $b^3$ \cite{Mukherjee2002} or $0.7b^3$ \cite{Meyers2009} where $b$ is the magnitude of the Burgers vector. In this formulation, $D_{\mathrm{GB}} \delta$ is the diffusion flux in the GB, and $\sigma \Omega$ is the work performed by the stress during an elementary diffusion jump \cite{Keblinski1998, Yamakov2002}. It should be noted that the original paper by Coble \cite{Coble1963} reports the value of $A_C$ for spherical grains as 148. However, we performed the derivation by Coble and found a value of $A_C = 46.347$. This agrees with the values of 47 reported by Keblinski \textit{et al.} \cite{Keblinski1998} and $148/\pi$ by Wang \textit{et al.} \cite{Wang2011}. The maximum difference between these different results for $A_C$ is only a factor of three; thus, it is effectively negligible compared to the orders of magnitude difference in creep rate resulting from, for example, $d$ in \cref{Eq:Coble}. In \cref{Sec:ParameterBased}, \cref{Eq:Coble} is used as a basis for the parameter-based Coble creep model.

Homogeneous grain elongation along the tensile stress direction during diffusional creep requires GB sliding as a geometrically necessary accommodation process \cite{Yamakov2002, Courtney2005}. In this context, GB sliding happens sequentially to diffusional creep on a local atom-by-atom basis and does not involve changes in the relative positions of the centers of mass of individual grains (i.e., their coordinates normalized by the simulation box dimensions are constant) \cite{Courtney2005}. If GB sliding were absent, voids/microcracks would form along the GBs. Since GB sliding and diffusional creep occur sequentially and sliding is faster than diffusion \cite{Courtney2005}, the net creep rate is the rate of the slower process, which is diffusional creep.

\subsection{Diffusional creep}

High-temperature deformation and grain growth are coupled phenomena. Deformation enhances grain growth, and, in turn, grain growth can either \textit{increase} or \textit{decrease} the deformation rate depending on the interplay between the GB processes governing this coupling, which include GB diffusion, GB sliding, GB migration, and grain rotation \cite{Haslam2004}. Although the Coble creep formula (\cref{Eq:Coble}) implies that grain growth reduces the creep rate, experimental evidence suggests that the strain rate can also increase during deformation due to GB migration, which acts as a stress-relaxation mechanism \cite{Haslam2004}. To suppress grain growth, an idealized nanocrystalline microstructure of uniform grain sizes and shapes is usually used in MD creep simulations \cite{Keblinski1998, Yamakov2002}. The geometry of grains can be closely modeled by the space-filling 14-sided polyhedron known as the truncated octahedron \cite{Yamakov2002, Olander2017}. Truncated-octahedral grains are nearly spherical, yet they display realistic grain features like triple lines and quadruple junctions. Atomsk \cite{Hirel2015} was used to generate 16 truncated-octahedral grains with random misorientations, and average grain diameters of 14, 16, 18, and 20 nm. Based on our tests, smaller grain sizes (10 and 12 nm) suffered inconsistent evolution in creep simulations at the highest temperature (i.e., 2000 K) and the lowest stress (i.e., 200 MPa). Random grain misorientations lead to a relatively small fraction of low-angle GBs \cite{Yamakov2002, Haslam2004}; thus, all GBs within a polycrystal can be considered high-angle GBs. Atoms with distances smaller than $0.45a$, where $a$ is the lattice parameter, were deleted to prevent unrealistically high forces as an artifact of the structure generation process.

Based on initial tests, we found that the Kocevski potential stabilizes voids along GBs in both bicrystals and polycrystals. This is likely due to the U-U repulsion in the Kocevski potential and its inability to simulate metallic U \cite{AbdulHameed2024}. Thus, the Kocevski potential cannot be used for the study of Coble creep phenomena. Additionally, governed by the migration of vacancies and atoms along GBs, accurate modeling of Coble creep requires reliable predictions of the energetics of vacancies and interstitials. The Tseplyaev potential provides accurate formation and migration energies for both stoichiometric and non-stoichiometric point defects that closely agree with DFT values \cite{Kuksin2016,Tseplyaev2016,AbdulHameed2024}. On the other hand, the Kocevski potential fails to reproduce accurate formation energies for U-rich and stoichiometric conditions \cite{AbdulHameed2024}. For these reasons, only the Tseplyaev potential is used to simulate the creep behavior of UN.

Supercells are equilibrated in the \textit{NPT} ensemble using the Nosé-Hoover thermostat and barostat at temperatures of 1700--2000 K and zero pressure for 200 ps. Then, tensile stresses in the range of 200--500 MPa are applied to the supercells in the $x$-direction for 5 ns. The calculations were repeated using three different initial velocity distributions.

Diffusional creep is typically observed for $T > 0.5 T_m$, where $T_m$ is the melting point \cite{Courtney2005}. For UN, the experimental melting temperature is about 3035 K at a nitrogen vapor pressure of 1 atm \cite{Hayes1990IV}, and the Tseplyaev potential predicts thermodynamic melting of UN at about 2700 K \cite{AbdulHameed2024}. Thus, diffusional creep is expected to be observed at the studied temperature range of 1700--2000 K. On the other hand, the applied stresses (i.e., 200--500 MPa) are an order of magnitude higher than typical experimental stresses at which creep is observed and will lead to strain rates in the range of $10^6$ s$^{-1}$, which is several orders of magnitude higher than experimental creep rate values. However, such high stresses are required for creep to be observed within MD time scales. If we decrease the applied stress by an order of magnitude and increase the grain size from 10 nm to 10 $\mu$m, the strain rate becomes $10^{-4}$--$10^{-5}$ s$^{-1}$---a typical experimental creep rate value. However, this would increase the computational cost to achieve a similar creep strain by approximately $10^4$ times. It is expected that the fundamental physics of Coble creep is captured, given the computational setup applied here, and thus the extrapolation of conclusions from high stress/small grain simulations to more realistic stresses and grain sizes is considered reasonable. Ultrahigh strain rates have been traditionally used in MD modeling of Coble creep and tensile testing \cite{Keblinski1998,Yamakov2002,Desai2008,Haslam2004,AbdulHameed2024b}.

To visualize the microstructural details of the supercells, the centrosymmetry parameter (CSP) and the dislocation extraction algorithm (DXA) are utilized. The CSP has been calculated by OVITO for snapshots of the simulated supercells using the minimum-weight matching algorithm \cite{Larsen2020}. The CSP quantifies lattice disorder and is defined as \cite{Kelchner1998}:
\begin{equation}
\mathrm{CSP} = \sum_{i=1}^{N/2} |\mathbf{r}_i + \mathbf{r}_{i+N/2}|^{2},
\end{equation}
where $N$ represents the number of nearest neighbors around an atom (for NaCl, $N = 6$), and $\mathbf{r}_i$ and $\mathbf{r}_{i+N/2}$ are vectors extending from the atom in question to a pair of opposite neighboring atoms. In a perfectly centrosymmetric crystal, contributions from all neighboring pairs cancel each other, yielding a CSP of zero. The CSP method is resilient to thermal noise; for a pristine crystal, it produces a single peak starting at zero, with its width broadening as temperature $T$ increases. A second peak in the CSP graph indicates a different local structure, such as dislocations or highly disordered grain boundaries (GBs) \cite{Bulatov2006, Larsen2020}. The CSP has units of squared distance, typically expressed as $a^2$, where $a$ is the lattice constant. The CSP serves as an indirect analysis method. For direct visualization of dislocations, OVITO's DXA \cite{Stukowski2012} is used, particularly applied to the U sub-lattice. A drawback of the DXA is its limitation in identifying predefined dislocations in single-component crystals, so dislocations with a Burgers vector outside the predefined families go undetected. This limits the algorithm's ability to identify GB dislocations, which typically have much shorter Burgers vectors than those in the lattice \cite{Cai2016}. Given that each sub-lattice in the B1 structure of UN forms a face-centered cubic (FCC) lattice, applying DXA to the U sub-lattice allows for precise dislocation identification and visualization.

\section{Results}

\subsection{Direct simulation of Coble creep}
\label{Sec:CreepResults}

\begin{figure}[h]
    \centering
    \includegraphics[width=0.45\textwidth]{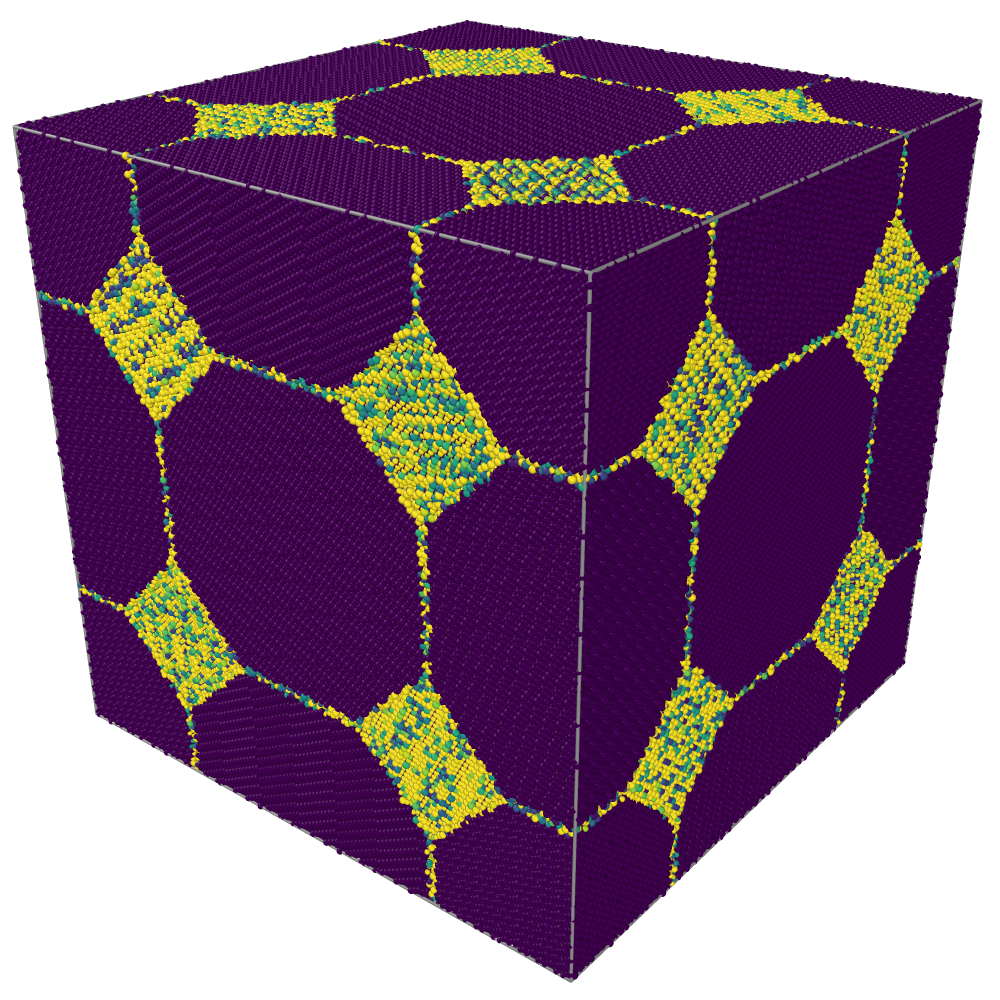}
    \caption{(Color online) UN polycrystalline supercell after energy minimization has been minimized at 0 K with a relative energy tolerance of $10^{-9}$. To visualize the GB structure, atoms have been colored in OVITO according to their centrosymmetry parameter (CSP), where dark purple atoms have low CSP and yellow atoms have high CSP. For better visibility, the maximum CSP has been set to 20 \AA$^{2}$.}
    \label{Fig:Structure}
\end{figure}

A model microstructure of UN whose energy has been minimized at 0 K with a relative energy tolerance of $10^{-9}$ is shown in \cref{Fig:Structure}. The numbers of atoms in the constructed polycrystals range between 1,547,344 atoms for the 14-nm polycrystal and 4,537,943 atoms for the 20-nm polycrystal, and the N/U ratio varies between 0.9999 and 1.0002. Each system is therefore either slightly hypo-stoichiometric or slightly hyper-stoichiometric, and any effect that may arise due to this slight off-stoichiometry is minor. Each creep rate is an average over three simulations using three different initial velocity distributions. To visualize the GB structure, atoms have been colored according to their CSP; yellow atoms have high CSP (i.e., high degree of miscoordination) and dark purple atoms have low CSP (i.e., low degree of miscoordination). Conceptually, low-angle GBs consist of isolated dislocation cores of high miscoordination that are separated by a nearly perfect crystal region of low miscoordination \cite{Keblinski1999}. On the other hand, high-angle GBs consist of overlapping dislocation cores and display a high degree of miscoordination that is uniformly distributed along the GB area \cite{Haslam2004}. Hence, it can be concluded from the figure that virtually all of the GBs in the supercell are high-angle GBs which are visualized as continuous uninterrupted yellow lines. This also applies to all the supercells employed in this work. Networks of GB dislocations have also been observed along the GBs using the DXA in OVITO. These GB dislocations are not preferentially aligned with any slip plane, are essentially immobile, and do not give rise to any dislocation-dominated deformation that would interfere with diffusional creep. Additionally, sustainable plastic deformation via the dislocation–nucleation mechanism is not possible because, e.g., the stress required to activate a Frank-Read source (a principal mechanism to sustain plastic deformation via dislocation multiplication \cite{Murty2013}) increases inversely with decreasing grain size ($\sigma \sim 1/d$), and the size of a Frank-Read source cannot exceed the grain size \cite{Yamakov2001, Yamakov2002}. Thus, our nanometer-sized grains are too small to accommodate a Frank–Read source that can be activated by typical stresses. It has also been visually confirmed that the microstructure does not undergo any micro-cracking, grain growth, GB sliding, or dislocation activity within grain interiors during the time scale of the simulation.

\begin{figure}[h!]
\centering
\begin{subfigure}{0.48\textwidth}
    \includegraphics[width=\textwidth]{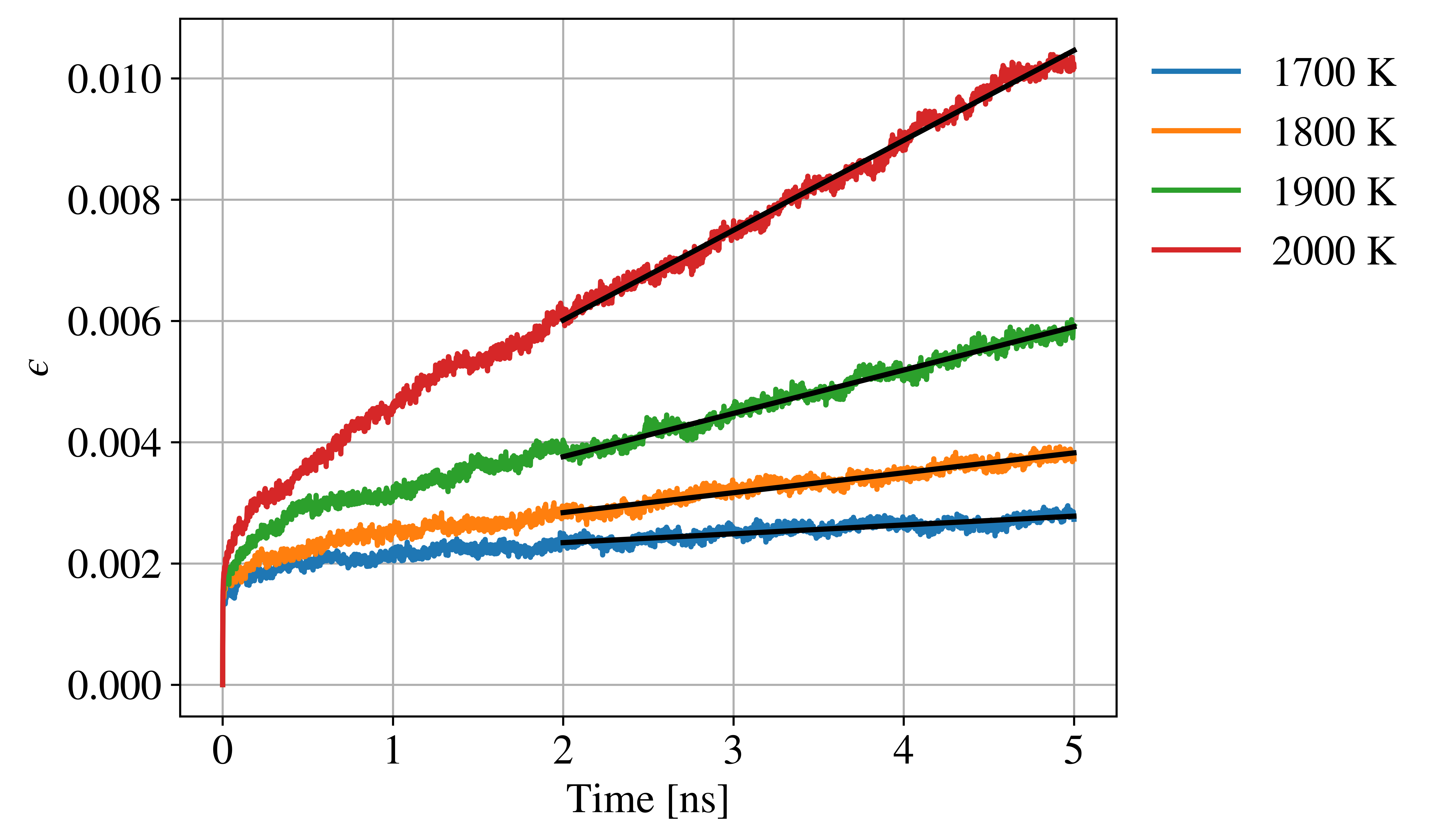}
    \caption{}
    \label{Fig:strain-vs-time-14-nm-200-MPa}
\end{subfigure}
\hfill
\begin{subfigure}{0.48\textwidth}
    \includegraphics[width=\textwidth]{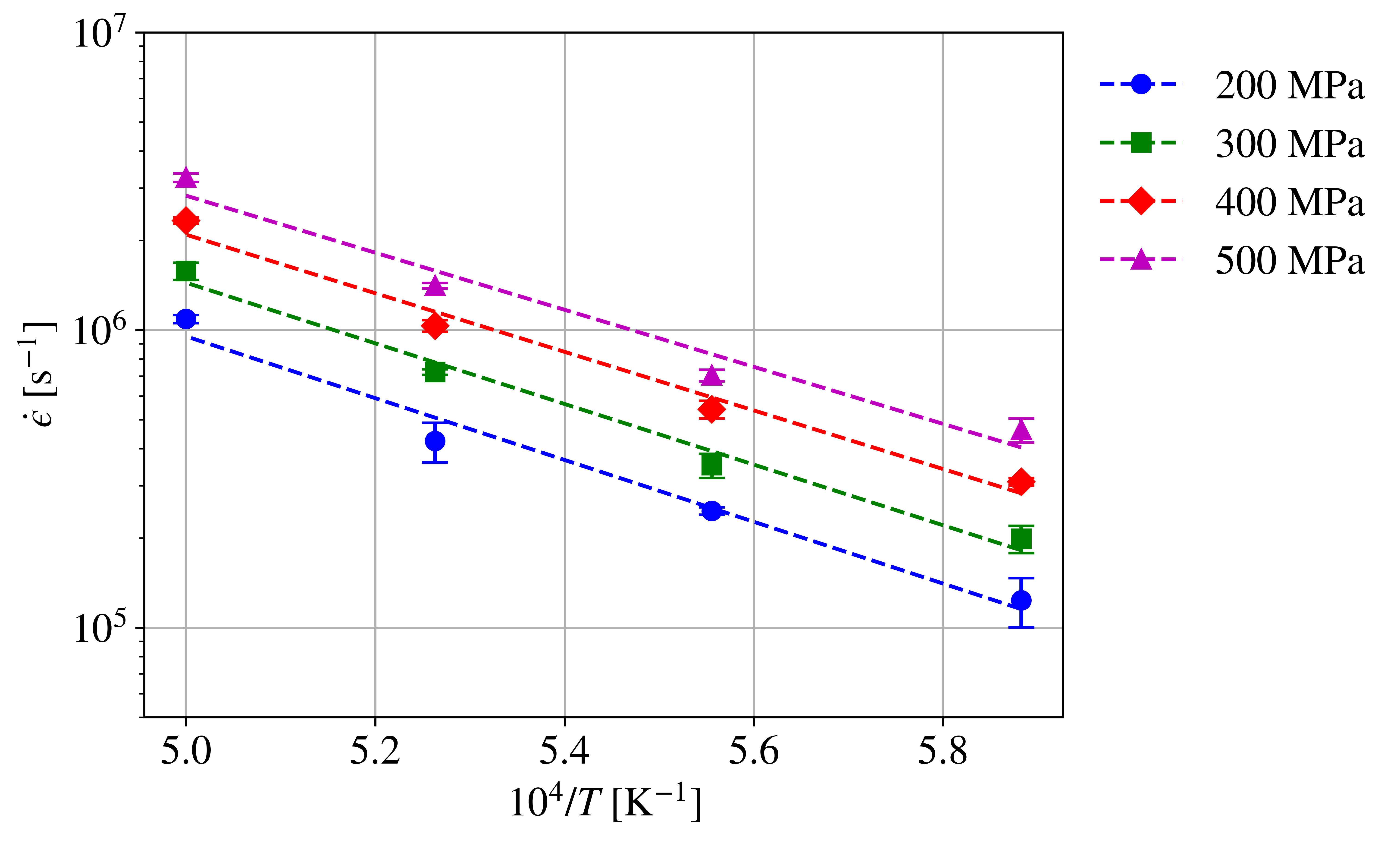}
    \caption{}
    \label{Fig:strain-rate-vs-temp-16-nm}
\end{subfigure}
\hfill
\begin{subfigure}{0.48\textwidth}
    \includegraphics[width=\textwidth]{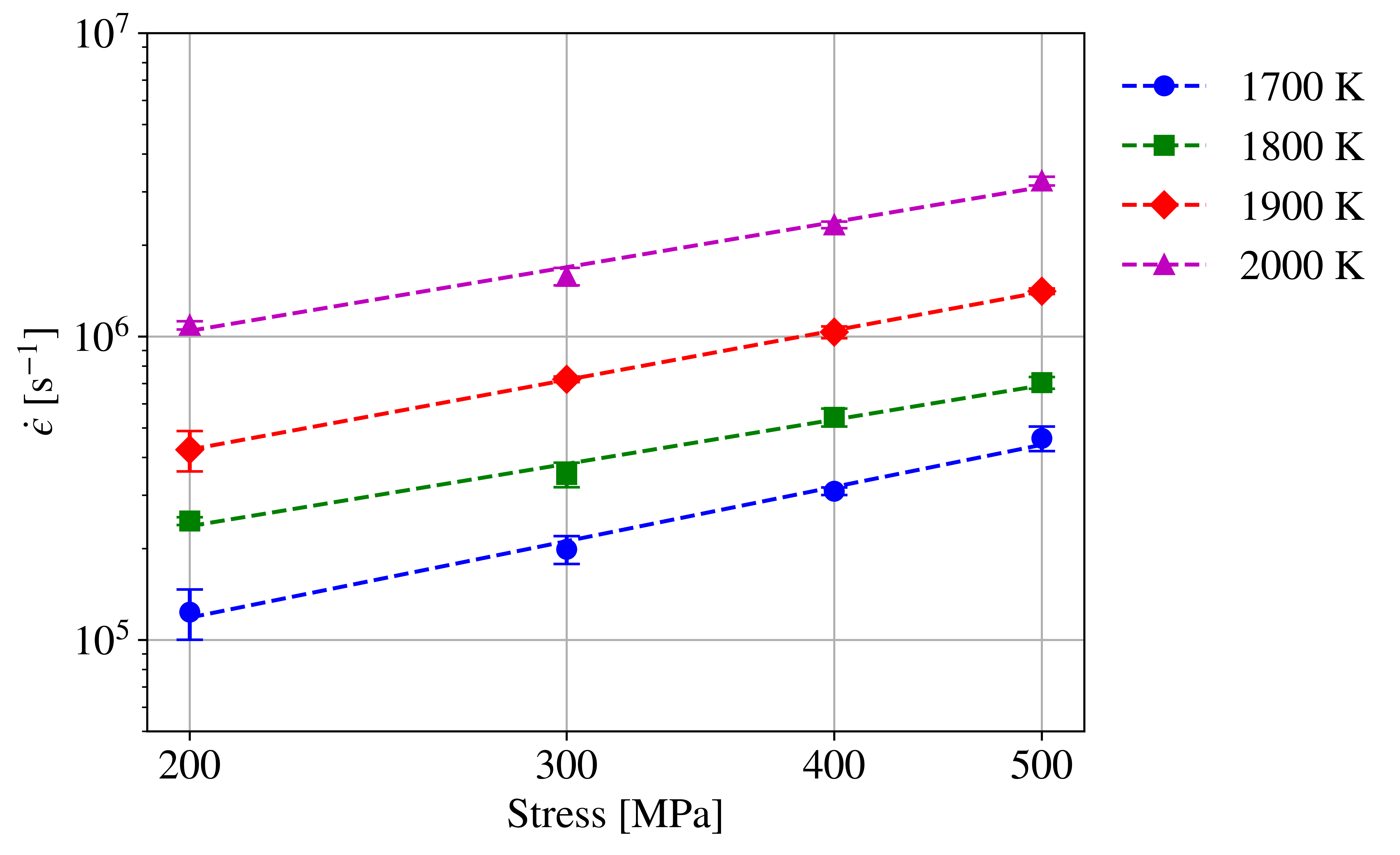}
    \caption{}
    \label{Fig:strain-rate-vs-stress-16-nm}
\end{subfigure}
\hfill
\begin{subfigure}{0.48\textwidth}
    \includegraphics[width=\textwidth]{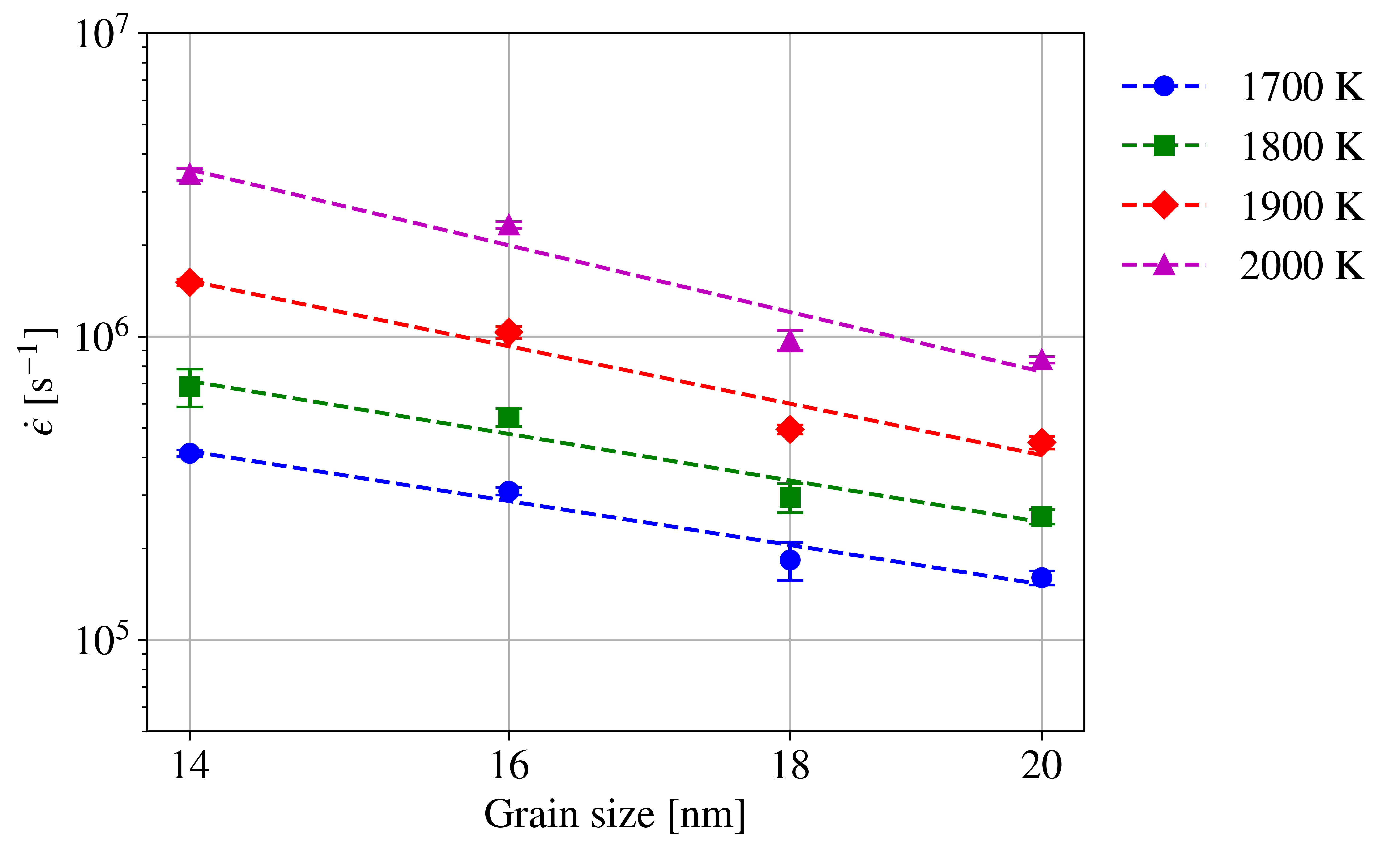}
    \caption{}
    \label{Fig:strain-rate-vs-grain-size-400-MPa}
\end{subfigure}
\caption{(Color online) \textbf{(a)} Variation of creep strain with time for a supercell with a grain size of 14 nm under constant stress of 200 MPa. \textbf{(b)} Semi-log plot of the variation of creep rate with inverse temperature for a supercell with a grain size of 16 nm. \textbf{(c)} Log-log plot of the variation of creep rate with stress for a supercell with a grain size of 16 nm. \textbf{(d)} Log-log plot of the variation of creep rate with grain size for supercells under constant stress of 400 MPa. In all figures, solid black lines correspond to the curve fits, and error bars correspond to one standard deviation.}
\label{Fig:Creep}
\end{figure}

Plots from the Coble creep simulations are shown in \cref{Fig:Creep}. An example of the strain variation with time is shown in \cref{Fig:strain-vs-time-14-nm-200-MPa}. After the instantaneous elastic strain, a transient strain stage is observed in which the strain rate decreases with time until it attains its steady-state value. Temperatures and stresses were chosen to maintain steady-state creep throughout the simulation while simultaneously ensuring a sufficient signal-to-noise ratio in the strain response. The strain rate is calculated using a linear fit to the final 3 ns of the strain versus time curve where steady-state creep is reached. The data was analyzed to parameterize \cref{Eq:MBDEq}. To calculate the activation energy of creep, the Arrhenius plots of the strain rate versus inverse temperature (e.g., \cref{Fig:strain-rate-vs-temp-16-nm}) have been fitted and an average value over all grain sizes and stresses calculated. As shown in \cref{Tab:CreepParams}, the calculated activation energy is $Q = 1.79 \pm 0.33$ eV, which is smaller than the activation energy of the Hayes correlation for thermal creep in UN (\cref{Eq:Hayes}, $Q = 3.39$ eV \cite{Hayes1990II}), which assumes a dislocation mechanism, and closer to the activation energy for GB diffusion in UN assumed by Konovalov \textit{et al.} (i.e., $Q$ = 2.65 eV) \cite{Konovalov2016}. This indicates that Coble creep can be a competitive mechanism to dislocation creep in UN. It should be noted, however, that the large uncertainty in the activation energy can lead to more than an order of magnitude uncertainty in the predicted creep rate.

\begin{table}[h]
\centering
\caption{Coble creep parameters estimated from fitting the simulation results of the diffusional creep of UN. Uncertainties represent one standard deviation.}
\footnotesize
\begin{tabular}{ccc}
\hline
Activation energy, $Q$ [eV] & Stress exponent, $n$ & Grain size exponent, $q$ \\
\hline
$1.79 \pm 0.33$             & $1.35 \pm 0.22$      & $3.58 \pm 0.87$ \\
\hline
\end{tabular}
\label{Tab:CreepParams}
\end{table}

To determine the dependence of the creep strain rate on the applied stress for each grain size and temperature, log-log plots of strain rate versus stress (\cref{Fig:strain-rate-vs-stress-16-nm}) have been fitted and the averaged value is shown in \cref{Tab:CreepParams}. A strain exponent $n = 1.35 \pm 0.22$ indicates that the stress dependence is nearly linear, which is characteristic of diffusional creep, and which agrees with the stress exponent found by Rogozkin \textit{et al.} \cite{Rogozkin2003}. Following the same procedure for the stress exponent, the grain size exponent has been determined (\cref{Fig:strain-rate-vs-grain-size-400-MPa}) for each applied stress and temperature, and the final averaged value is also shown in \cref{Tab:CreepParams}. It can be seen that the average grain size exponent $q = 3.58 \pm 0.87$ is very close to the grain size dependence of Coble creep ($q = 3$), but with relatively high uncertainty ($\sim$24\%). Uncertainties of the fitting parameters in \cref{Tab:CreepParams} are estimated by dividing the standard deviations by the mean values. It should be emphasized that only a subset of the data is shown in \Cref{Fig:Creep}, but the entire data set for all stresses, temperatures, and grain size are utilized in the determination of the average values in \cref{Tab:CreepParams}.

To confirm that the deformation mechanism is homogeneous (i.e., the total deformation of the supercell is proportional to the deformation of the individual grains), we visually confirmed that the positions of the centers of mass of the grains, normalized by the instantaneous lengths of the cubic simulation box, did not move during the time scale of the simulation. Additionally, using the ``displacement vectors'' modifier in OVITO \cite{Stukowski2010} it was confirmed that only atoms along the GBs undergo motion during the time scale of the simulations. That is, only Coble creep is observed in our simulations and Nabarro-Herring creep is essentially absent at the considered temperatures and time scale.

\subsection{Parameter-based Coble creep model}
\label{Sec:ParameterBased}

Following the methodology of Cooper \textit{et al.} \cite{Cooper2021} for U$_3$Si$_2$ and Galvin \textit{et al.} \cite{Galvin2025} for UO$_2$, we also construct a parameter-based Coble creep model. In this approach, the parameters of the Coble creep formula (\cref{Eq:Coble}) are calculated independently. 

\subsubsection{GB diffusivity}
\label{Sec:GBD}


In ionic compounds like UN, the diffusional process requires ambipolar coupling to avoid compound decomposition \cite{Gordon1974}. This means that U and N atoms must diffuse in stoichiometric ratios with an effective diffusivity that is limited by the diffusivity of the slower species, i.e., uranium, along its fastest path. The total effective diffusivity is calculated based on Gordon's formula \cite{Gordon1974} for ambipolar diffusion in an ionic solid A$_{\alpha}$B$_{\beta}$ (For UN, $\alpha = \beta = 1$):
\begin{equation}
D_\mathrm{eff} = \frac{(\alpha + \beta) D_\mathrm{A} D_\mathrm{B} }{\beta D_\mathrm{A} + \alpha D_\mathrm{B}}.
\label{Eq:Deff}
\end{equation}
\cref{Eq:Deff} differs from the usual formula for effective $D$ by including the extra term $(\alpha + \beta)$, which appears because Gordon's formulation is based on the molecular volume of A$_{\alpha}$B$_{\beta}$ whereas the traditional Coble creep formula (\cref{Eq:Coble}) includes the average atomic volume \cite{Kizilyalli1999, Desai2008}. Because we are dealing with polycrystals, it is difficult to exclusively track the atomic movement within the GBs to determine the GB diffusivity. Instead, we track the movement of all atoms within the polycrystals. In this picture, the total diffusivity of either U or N atoms has both bulk and GB components. According to the Hart formulation \cite{Hart1957, Riet2021}, the total diffusivity in a polycrystal can be divided between bulk and GB components according to:
\begin{equation}
D = c D_\mathrm{GB} + (1-c) D_b,
\end{equation}
where $c$ is the fraction of atoms that belong to the GBs ($c$ will be defined quantitatively later in \cref{sec:GB_width}), $D_\mathrm{GB}$ is the GB diffusivity, and $D_b$ is the bulk diffusivity. Thus, with the knowledge of the total diffusivity in the polycrystals and the bulk diffusivity, the GB diffusivity can be readily estimated. 

\begin{figure}[h!]
\centering
\begin{subfigure}{0.48\textwidth}
    \includegraphics[width=\textwidth]{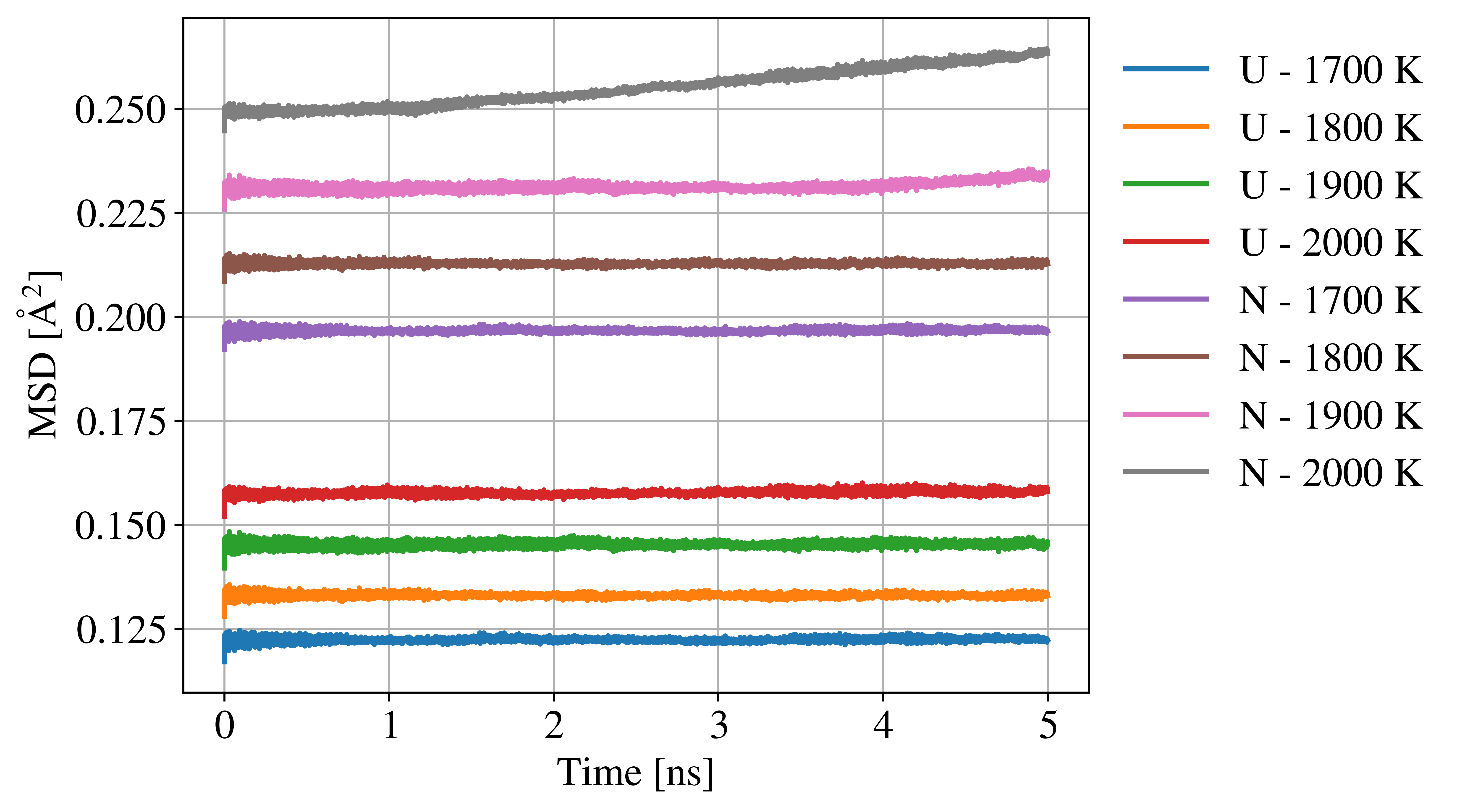}
    \caption{Bulk diffusivity (MSD)}
    \label{Fig:LatDiff}
\end{subfigure}
\hfill
\begin{subfigure}{0.48\textwidth}
    \includegraphics[width=\textwidth]{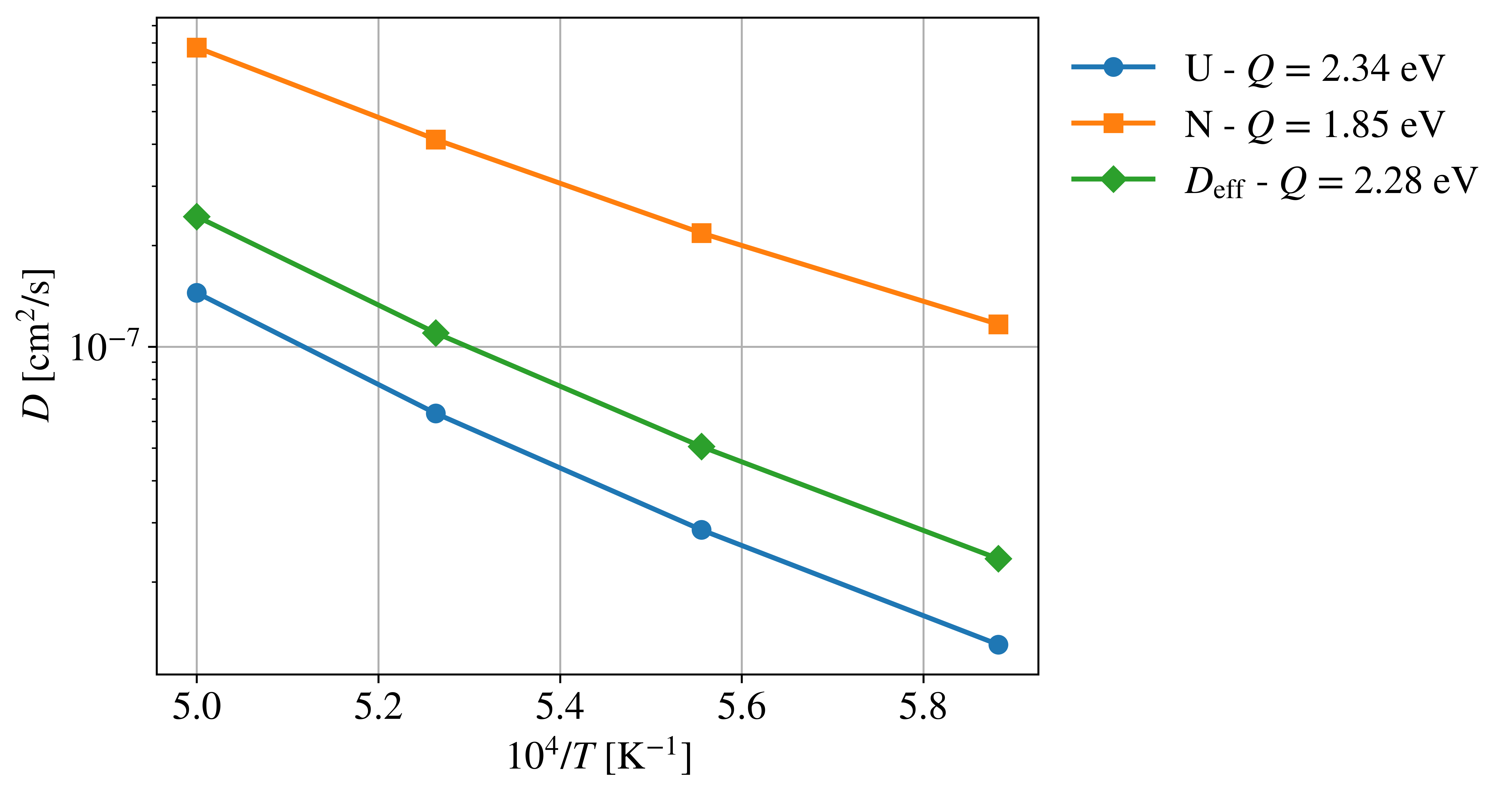}
    \caption{GB diffusivity (Arrhenius plot)}
    \label{Fig:GBDiff}
\end{subfigure}
\caption{(Color online) \textbf{(a)} Mean squared displacements (MSD) of U and N atoms within bulk UN at different temperatures. A fit of the MSD curve for N atoms at 2000 K over the final 3 ns gives $D = 5.91 \times 10^{-11}$ cm$^2$/s. \textbf{(b)} Arrhenius plot of the GB diffusivity of U and N atoms, as well as the effective GB diffusivity. Note that the standard deviation of $Q_\mathrm{U}$ is $\pm 0.03$ eV, and that of $Q_\mathrm{N}$ is $\pm 0.02$ eV.}
\label{Fig:Diff}
\end{figure}

To calculate the bulk diffusivities of U and N atoms, an unbound Schottky pair is randomly introduced within a $50 \times 50 \times 50$ UN supercell which is then equilibrated at temperatures of 1700--2000 K and zero pressure under the \textit{NPT} ensemble for 100 ps \cite{Arima2010}. After equilibration, the mean squared displacements (MSD) of both species are averaged every 1 ps for a total simulation time of 5 ns. The results are shown in \cref{Fig:LatDiff}. It is apparent that the U atoms do not diffuse within bulk UN and only undergo vibrational motion in their positions. N atoms start to show non-zero diffusivity only at 2000 K and only vibrate around their equilibrium positions at lower temperatures. According to the Einstein relation, $\mathrm{MSD} = 6 D t$ in the limit of an infinite time. Fitting the MSD curve for N atoms at 2000 K over the final 3 ns gives $D_b^\mathrm{N} = 5.91 \times 10^{-11}$ cm$^2$/s. This diffusivity value is negligible compared to the GB diffusivities of either U or N atoms (\cref{Fig:GBDiff}, which will be described later in this subsection). Because U atoms do not undergo any measurable diffusivity, we can conclude that bulk diffusivity is essentially non-existent and all the atomic movement in the polycrystals can be solely attributed to GB diffusion. 


The motion of all atoms in stress-free polycrystals with a grain size of 20 nm has been tracked. The polycrystal is equilibrated at the target temperature and zero pressure within the \textit{NPT} ensemble for 50 ps. Then, the MSD of all atoms of both types is averaged every 1 ps for a total averaging time of 5 ns. It was shown earlier that bulk diffusion is absent and all the atomic movement is attributed to diffusion along the GBs. Diffusion in the direction perpendicular to a GB is usually much smaller than the other two directions spanned by the GB, and GB diffusion is treated as a two-dimensional (2D) phenomenon \cite{Riet2021}. Thus, we calculate the GB diffusivity of each atomic species by a linear fit of the MSD of that species over the last 3 ns to the 2D Einstein relation:
\begin{equation}
D_\mathrm{GB}^\mathrm{U, N} = \frac{1}{c} \frac{\mathrm{MSD^\mathrm{U, N}}}{4 t}.
\label{Eq:GBD}
\end{equation}

The results of such calculations are shown in \cref{Fig:GBDiff}. The calculation of the variable $c$ for each temperature and grain size will be shown later. After getting the GB diffusivities of U and N atoms, the effective diffusivity is calculated based on Gordon's formula (\cref{Eq:Deff}). The effective activation energy of GB diffusion is $Q$ = 2.28 $\pm$ 0.09 eV and the prefactor is $D_0$ = 0.130 $\pm$ 0.075 cm$^2$/s (which has the typical order of magnitude). It can be seen that the activation energy of GB diffusion in stress-free polycrystals ($Q_{\mathrm{eff}} = 2.28$ eV) is closer to the assumed activation energy of GB diffusion in UN ($Q_\mathrm{GB}$ = 2.65 eV \cite{Konovalov2016}) than the fitted activation energy of Coble creep simulation ($Q$ = 1.79 eV). Additionally, we repeated the GB diffusion calculation for grains with sizes of 14--18 nm and found that the standard deviation of $Q_\mathrm{U}$ is $\pm$ 0.03 eV, and that of $Q_\mathrm{N}$ is $\pm$ 0.02 eV. Compared to this, the fitted activation energy of Coble creep has a larger uncertainty of $\pm$ 0.33 eV. This discrepancy can be attributed to the stress effect on GB diffusion. In general, GB diffusivity increases with the applied stress \cite{Haslam2004}, which might be manifested by a reduction in the activation energy. A reduction in the activation energy stems from an increase of the GB diffusivity at the higher temperatures for which the stress effect on the GB diffusivity is more pronounced. It is also observed that the activation energy for the effective diffusivity is nearly dominated by that of U atoms, which confirms the assertion that the GB diffusion of the slowest-moving species, i.e., U atoms, is rate-limiting for Coble creep. 

\subsubsection{Grain-boundary width}
\label{sec:GB_width}

Next, we closely analyze the GB width. At first, we need to differentiate between two concepts of GB width in high-purity materials \cite{Keblinski1999, Prokoshkina2013}: (\textit{a}) diffusional GB width, $\delta_d$, which corresponds to the width within which atoms diffuse, and (\textit{b}) structural GB width, $\delta_s$, which corresponds to the width within which miscoordinated atoms exist, whether they diffuse or not. Keblinski \textit{et al.} \cite{Keblinski1999} found in their atomistic simulations that $\delta_d$ is thermally activated and shows an Arrhenius dependence on temperature. However, their analysis was based on bicrystals, which usually contain symmetric tilt GBs and are not necessarily representative of realistic microstructures. Additionally, the analysis was conducted at temperatures near the melting point, which are not usually accessible in GB diffusion experiments due to the simultaneous contribution of bulk diffusion at these high temperatures. Thus, their results could not be experimentally verified. On the other hand, Prokoshkina \textit{et al.} \cite{Prokoshkina2013} conducted an extensive analysis of the measured GB widths of various metals and compounds and found that the GB widths measured in GB diffusion measurements do not show any dependence on temperature or grain size. The purpose of this section is to resolve this apparent conflict between atomistic simulations and experiments and provide a consistent definition for GB width that can be extracted from atomistic simulations and agrees with the experimental data. We begin by using our GB diffusivity data to extend the analysis from Keblinski \textit{et al.} to polycrystals of uniform size and shape and temperatures relatively far from the melting point.

The diffusion flux at the GB (in m$^3$/s) is a product of the GB diffusivity and diffusional GB width: $D_{\mathrm{GB}} \delta_d$, and can be expressed as follows:
\begin{equation}
D_{\mathrm{GB}} \delta_d = \lim_{t \rightarrow \infty} \frac{\Omega}{A} \frac{\sum_{i=1}^{N_{\mathrm{GB}}} |\Delta \mathbf{r}_i(t)|^2}{6t} \approx \lim_{t \rightarrow \infty} \frac{\mathrm{MSD}_{\mathrm{GB}}(T)}{6t} \frac{\Omega N_{\mathrm{GB}}(T) }{A},
\label{Eq:GBDiffFlux}
\end{equation}
where both the MSD of GB atoms, MSD${_{\text{GB}}}$, and the number of atoms \textit{diffusing} within the GB, $N_{\mathrm{GB}}$, are functions of temperature. In \ref{app}, we derive the following formula for the diffusional GB width of polycrystals:
\begin{equation}
\delta_d = \eta c V^{1/3},
\label{Eq:delta}
\end{equation}
where $V$ is the volume of a space-filling grain and $c$ is defined as the fraction of atoms that have undergone non-affine displacements of at least the nearest-neighbor distance in the limit of infinite simulation time. For UN, the nearest-neighbor distance is $0.5a$, where $a$ is the lattice parameter. $c$ can be estimated using the displacement-vectors (DV) modifier in OVITO \cite{Stukowski2010} applied to snapshots of the polycrystal. $\eta$ is a shape factor that depends on the grain shape. For truncated-octahedral grains, $\eta = 0.376$ (\cref{eq:shape}). Note that, unlike the calculation of the GB diffusivity which is based on averages taken during the MD simulation, the estimation of the GB width is a post-processing step.

Although \cref{Eq:delta} is derived for grains of uniform size and shape, it is principally applicable to grains of random sizes and shapes provided that the average \textit{grain volume} is treated as the volume of a truncated octahedral grain and the grain size distribution is relatively narrow. In this case, \cref{Eq:delta} can be readily applied to estimate $\delta_d$ for any polycrystal using $\eta = 0.376$. The advantage of this method is that it enables us to determine $\delta_d$ for a microstructure that contains realistic grain features like triple lines and quadruple junctions. In contrast to other methods based on bicrystal simulations, our proposed method simultaneously samples many GBs with various misorientations and gives an effective value for $\delta_d$ averaged over all of them.

\begin{figure}[h!]
\centering
\begin{subfigure}{0.48\textwidth}
    \includegraphics[width=\textwidth]{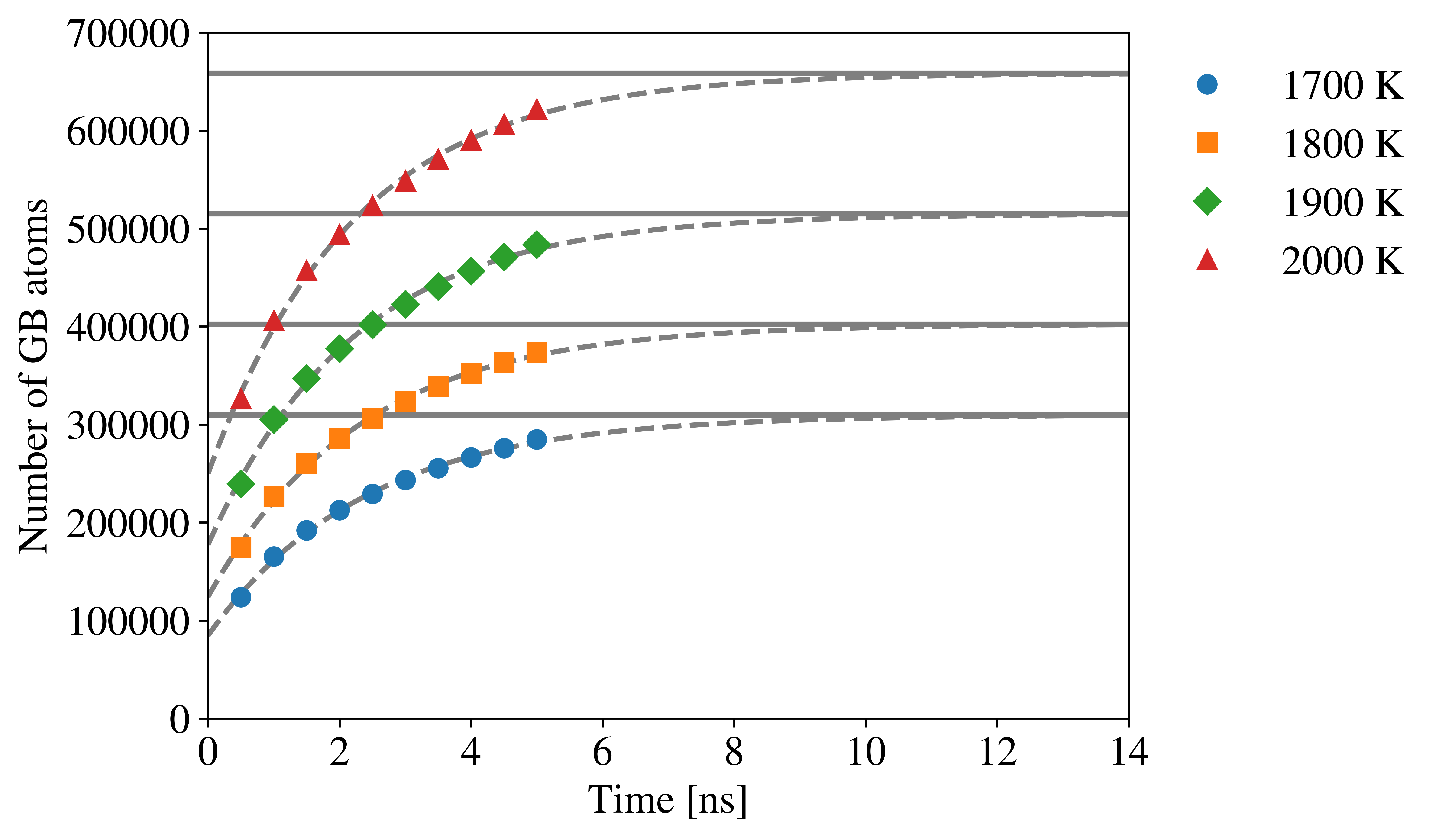}
    \caption{}
    \label{Fig:NGB}
\end{subfigure}
\hfill
\begin{subfigure}{0.48\textwidth}
    \includegraphics[width=\textwidth]{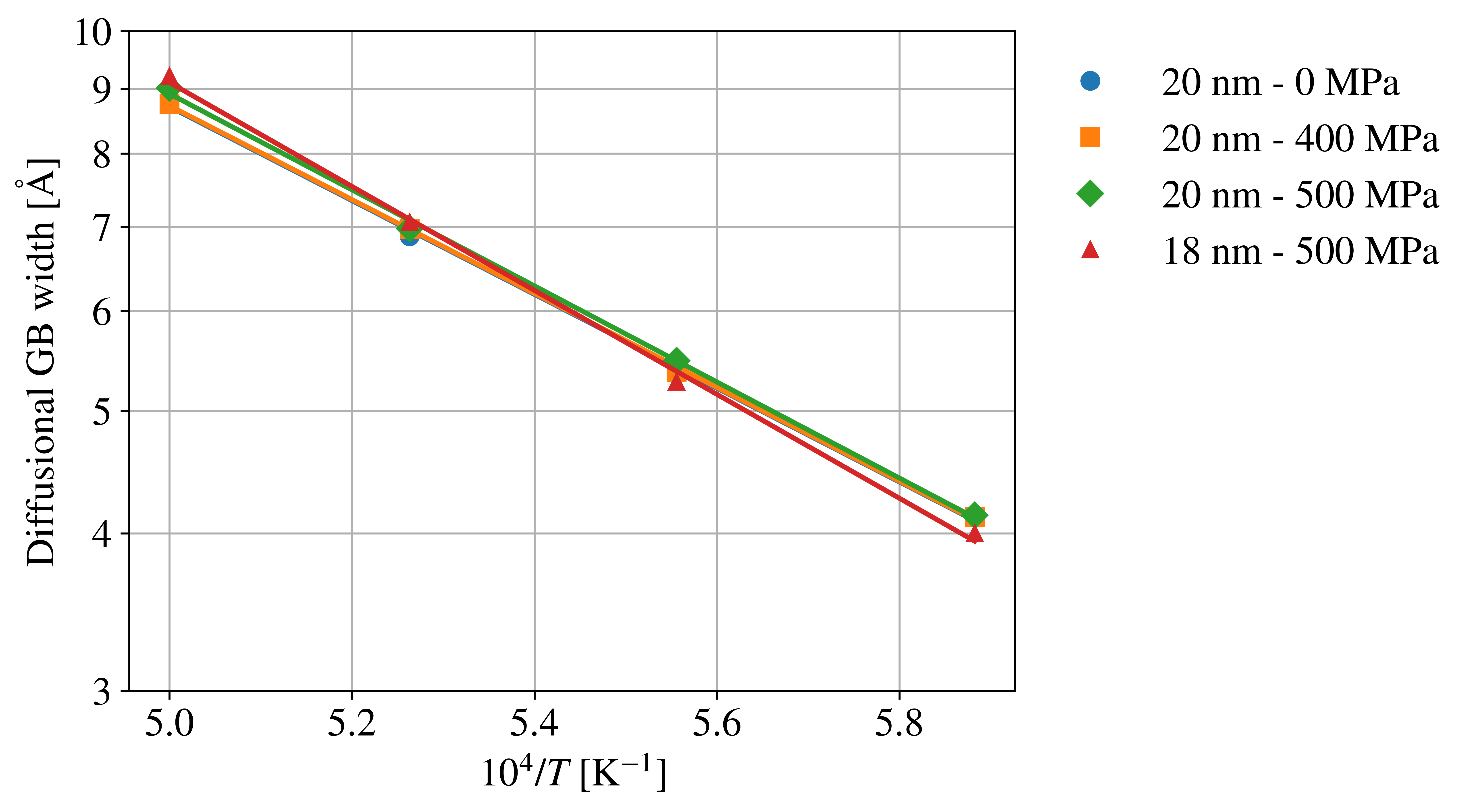}
    \caption{}
    \label{Fig:GBWidth}
\end{subfigure}
\caption{(Color online) \textbf{(a)} The variation of the number of GB atoms, $N_\mathrm{GB}$, with time for the polycrystal with a grain size of 20 nm under no stress. In the limit of infinite time, $N_\mathrm{GB}$ reaches a saturation value that increases with increasing temperature. \textbf{(b)} Arrhenius plot of the diffusional GB width for the tracked supercells. The activation energy of $\delta_d$ ranged between 0.74 and 0.82 eV.}
\end{figure}

This analysis is implicitly based on two assumptions. First, in the limit of a very long time, the number of GB atoms that have moved by at least the nearest-neighbor distance will saturate to a value determined by the temperature. This was assumed by Keblinski \textit{et al.} \cite{Keblinski1999} but never proved. To demonstrate that this is indeed the case, we use OVITO's DV modifier to track the number of diffusing atoms in a stress-free 20-nm supercell. To probe the effect of the stress and the grain size on the diffusional GB width, we conduct the same analysis for the 20-nm supercell under stresses of 400 MPa and 500 MPa and the 18-nm supercell under a stress of 500 MPa. In all cases, the number of diffusing atoms is fitted to a saturation function of the form:
\begin{equation}
N_\mathrm{GB} = N_0 \left( 1- e^{- \lambda t} \right) + C.
\end{equation}
In the limit of infinite time, the exponential approaches zero, and $N_\mathrm{GB} = N_0 + C$. Note that $N_\text{GB}$ represents the asymptotic (saturation) number of atoms participating in Coble creep once steady-state grain boundary diffusion is fully established. The result of the fitting for the 20-nm supercell under no stress is shown in \cref{Fig:NGB}, which demonstrates that $N_{\mathrm{GB}}$ perfectly follows a saturation function, and that the saturated value of $N_{\mathrm{GB}}$ increases with increasing temperature. Similar trends have been observed for the other tracked supercells. The saturated values of $N_\mathrm{GB} = N_0 + C$ are then fitted to an Arrhenius-type function. The results of such fitting for the tracked supercells are shown in \cref{Fig:GBWidth}, which shows that the diffusional GB width perfectly follows an Arrhenius behavior that is independent of supercell grain size and applied stress. The activation energy for all tracked supercells is 0.74 $\pm$ 0.01 eV for the 20-nm supercell at all stress values and 0.82 eV for the 18-nm supercell. While the effect of grain size on $Q_\delta$ is quite small, the effect of stress is essentially non-existent. Based on their bicrystal calculation, Keblinski \textit{et al.} \cite{Keblinski1999} estimated the activation energy of the GB width in Pd to be 0.22 eV.

While these results corroborate the computational findings of Keblinski \textit{et al.}, this Arrhenius dependence still contradicts the experimental observations that the measured GB width is independent of temperature. This leads to our second assumption: the GB width in the Coble creep formula corresponds to a temperature-independent (effective) GB width, which is calculated by extrapolating the diffusional GB width to the melting point where all miscoordinated atoms are assumed to have undergone diffusion. Extrapolating the Arrhenius curves in \cref{Fig:GBWidth} to 2700 K (the melting point as predicted by the Tseplyaev potential \cite{AbdulHameed2024}), the effective GB width is found to be $\delta = 2.69 \pm 0.08$ nm. This GB width should correspond to the diffusion of all the GB atoms and is the one we use in the parameter-based Coble creep formulation. We show later that this value of the GB width provides an excellent prediction of the prefactor in the phenomenological Coble creep formula, which supports our second assumption.

Lastly, we need to discuss the $c$ parameter in \cref{Eq:delta}, which is used in \cref{Eq:GBD} to estimate the GB diffusivity. In analogy with the two definitions of the GB width, there are two definitions of the $c$ parameter: (\textit{a}) the fraction of atoms that diffuse within the GBs, and (\textit{b}) the fraction of miscoordinated atoms existing within the GBs, whether they diffuse or not. A third definition \cite{Konovalov2016} is to estimate $c$ as the volume fraction of the GBs based on the formula:
\begin{equation}
c = \frac{3 \delta}{d}
\label{Eq:Vol}
\end{equation}
where $\delta$ is the GB width, and $d$ is the grain diameter. The same procedure of calculating the GB width is used to extract the $c$ parameter of the 20-nm polycrystal to calculate the GB diffusivity in \cref{Sec:GBD}. Note that the \textit{saturated} number of GB atoms is linearly proportional to the diffusional GB width (\cref{Eq:dNGB}), and, thus, both follow an Arrhenius behavior with the same activation energy. Like the GB width, we extrapolate the Arrhenius plot of the saturated number of GB atoms up to the melting temperature to extract the number of miscoordinated atoms. This number is then divided by the total number of atoms in the supercell to get $c$. Based on this procedure, $c$ = 0.44 $\pm$ 0.01 for the stress-free 20-nm polycrystal. It should be noted that if we used $\delta$ = 2.69 nm, and $d$ = 20 nm in \cref{Eq:Vol}, we get $c$ = 0.40. That is, the definitions of $c$ based on the volume fraction or the atomic fraction give very close results and can be used interchangeably. For macroscopic systems with $d$ in the $\mu$m-range, $c \sim 10^{-3}$. In effect, $c$ penalizes the nm-sized structures for having grains with a large surface-to-volume ratio.

\subsubsection{Vacancy volume}

The remaining unknown parameter in the Coble creep model (\cref{Eq:Coble}) is the vacancy volume. To calculate the vacancy volume in UN, we construct $6 \times 6 \times 6$ perfect supercells and calculate the volume of the minimized structure at 0 K. Then, we separately delete U and N atoms from the center of the supercell, minimize the energy of the defective structure, and finally calculate the vacancy volume as the volume difference between the pristine and defective crystals. All structures have been minimized with a relative energy tolerance of 10$^{-9}$. The result of such calculations is presented in \cref{Tab:VacancyVolume}, where the atomic volume is also given for reference. We use the values of the Tspelyaev potential, but those calculated by the Kocevski potential are also shown for comparison. It can be observed that the Kocevski potential predicts a positive volume for the N vacancy, which is unphysical. The reason for this is that upon removing the N atom, there is no longer a screening/shielding between the U atoms, which repel each other and lead to an increase in the volume of the supercell with the vacancy. This is due to the inability of the Kocevski potential to model metallic U. From \cref{Tab:VacancyVolume}, we take the absolute value of $\text{V}_\text{U}$ as the vacancy volume, i.e., $\Omega$ = 4.963 \AA$^3$. The vacancy volume is nearly one-third of the atomic volume (13.909 {\AA$^3$}), indicating that the commonly used approximation of equating $\Omega$ to atomic volume \cite{Courtney2005} is not accurate for UN. The discrepancy arises from the local lattice relaxation and volume shrinkage induced by the formation of a vacancy, highlighting the importance of calculating $\Omega$ explicitly.
To conclude, all the values for the parameter-based Coble creep model are summarized in \cref{Tab:Params}.

\begin{table}[h!]
\centering
\caption{Vacancy and atomic volumes (in \AA$^3$) as predicted by the Tseplyaev and Kocevski potentials. $\Omega(\mathrm{atom})$ is the average volume of a single atom.}
\footnotesize
\begin{tabular}{cccc}
\hline
                    & $\Omega(\mathrm{V}_\mathrm{U})$ & $\Omega(\mathrm{V}_\mathrm{N})$  & $\Omega(\mathrm{atom})$ \\
\hline
Tseplyaev potential & $-4.963$      & $-0.227$       & 13.909 \\
Kocevski potential  & $-1.990$      & 4.350          & 14.685 \\
\hline
\end{tabular}
\label{Tab:VacancyVolume}
\end{table}

As a sanity check of our calculated parameters, according to the Mukherjee-Bird-Dorn phenomenological model of Coble creep \cite{Mukherjee2002}, the product $A_C \Omega \delta$ should numerically equal $50b^4$ where $b$ is the Burgers vector. With $A_C$ = 46.347, we find that $A_C \Omega \delta$ = 6195 \AA$^4$, whereas $50b^4$ = 6691 \AA$^4$, with a difference of only 7.52\%. This also supports our method in calculating the GB width and its associated assumptions. In other words, although atomistic simulations reveal a temperature-dependent diffusional GB width, this behavior is not attributed to the limited timescales accessible to MD. Rather, our analysis shows that the number of diffusing atoms within GBs saturates with time and increases with temperature, resulting in an Arrhenius dependence of the diffusional GB width. In contrast, experimental GB widths reflect a long-time effective value that is typically temperature-independent. Additionally, the experimentally measured GB width can vary depending on the measurement technique. High-resolution transmission electron microscopy (HRTEM) yields the structural GB width, which is precise but not directly relevant to diffusional creep studies \cite{Prokoshkina2013}. In contrast, GB widths inferred from diffusion data are more relevant but are less reliable due to uncertainties associated with data fitting \cite{Prokoshkina2013}. To reconcile simulation and experimental perspectives, we extrapolated the temperature-dependent diffusional GB width to the melting point and defined this extrapolated value as the effective GB width to be used in the Coble creep formulation. This value (2.69 nm) reproduces the expected prefactor in the Coble creep equation and provides a consistent physical basis for incorporating MD-based GB width estimates into mesoscale models. It should be mentioned that the proposed method for estimating the effective GB width is general and transferable, as it relies solely on atomistic trajectories and makes no assumptions about the material's chemistry, crystal structure, or defect energetics.

\begin{table}[h!]
\centering
\caption{Values for the parameter-based Coble creep model.}
\footnotesize
\begin{tabular}{lc}
\hline
Coble prefactor, $A_C$ & 46.347 \\
Vacancy volume, $\Omega$ & $4.963 \times 10^{-30}$ m$^3$ \\
Effective GB width, $\delta$ & $2.69 \times 10^{-9}$ m \\
GB diffusivity prefactor, $D_0$ & $1.30 \times 10^{-5}$ m$^2$/s \\
GB diffusivity activation energy, $Q$ & 2.28 eV \\
\hline
\end{tabular}
\label{Tab:Params}
\end{table}

\subsubsection{Comparison of Creep Predictions}

\begin{figure}[h!]
    \centering
    \includegraphics[width=0.5\textwidth]{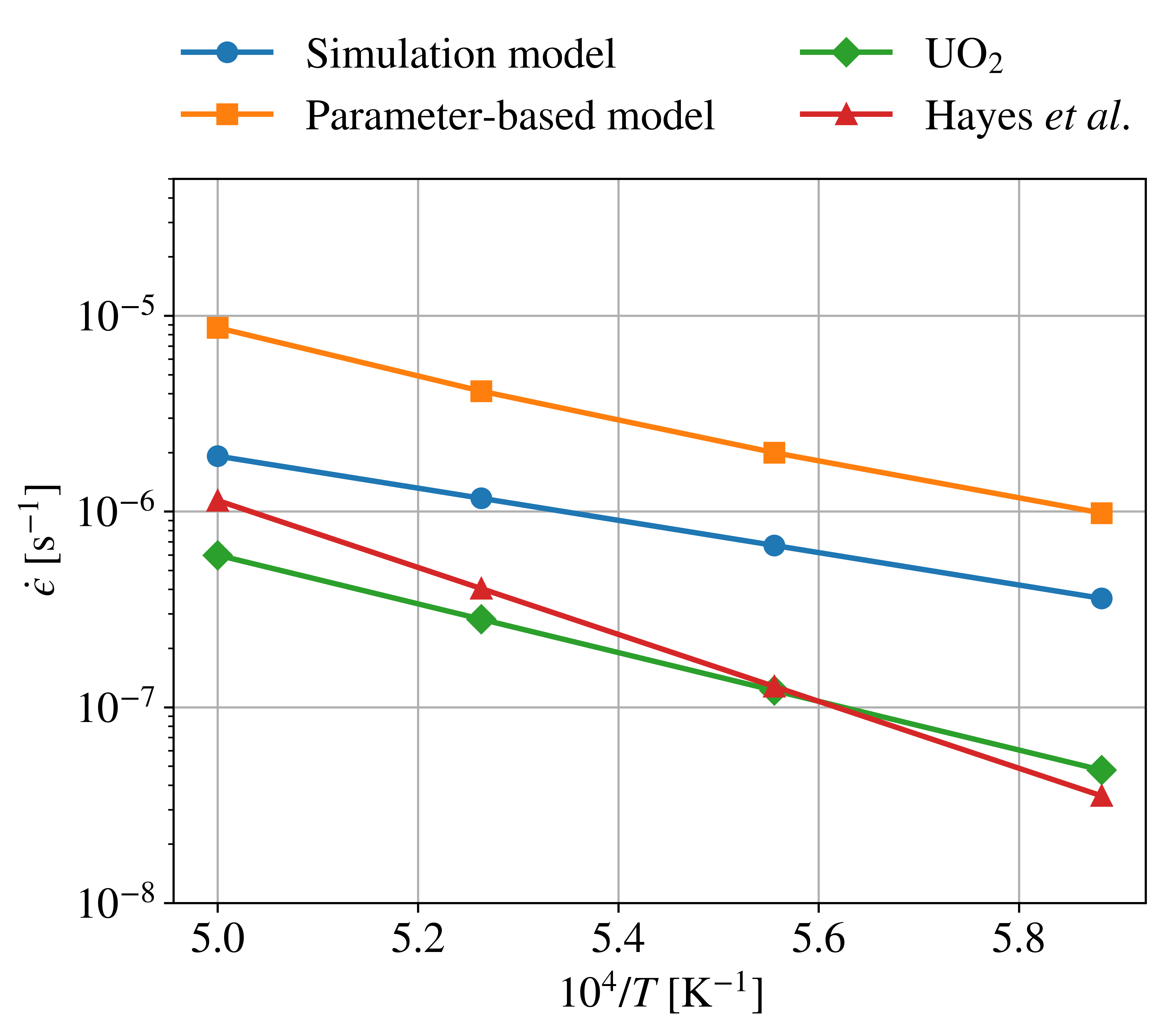}
    \caption{Comparison between the simulation and parameter-based models of Coble creep in UN developed in this work to Coble creep in UO$_2$ in the temperature range 1700--2000 K for a stress of 15 MPa, and a grain size of 10 $\mu$m. The UO$_2$ data is taken from \cite{Galvin2025}. The Hayes \textit{et al.} \cite{Hayes1990II} correlation for dislocation creep is also presented for comparison.}
    \label{Fig:UNvsUO2}
\end{figure}

We compare the simulation and parameter-based models of Coble creep in UN developed in this work to Coble creep in UO$_2$ as calculated by Galvin \textit{et al.} \cite{Galvin2025}. The prefactor in the simulation model was calculated by utilizing the values in \cref{Tab:CreepParams} within \cref{Eq:MBDEq}. The prefactor was found to have a numerical value of $A = 3.424 \times 10^{-26}$. For both fuels, a stress of 15 MPa, and a grain size of 10 $\mu$m were assumed. The result of such a comparison is shown in \cref{Fig:UNvsUO2}. It can be seen that the two models of Coble creep in UN agree to less than an order of magnitude, especially at lower temperatures. Additionally, irrespective of the creep model, the Coble creep rate of UN is approximately one order of magnitude larger than that of UO$_2$. It is also an order of magnitude larger than the Hayes \textit{et al.} correlation for dislocation creep in UN. Although a stress of 15 MPa is outside the stress range of the Hayes creep correlation (i.e., 20--34 MPa), it is not very far from the range's lower limit, and we assume it can at least provide an estimate of the dislocation creep rate for stresses slightly smaller or larger than the reported stress applicability range. This comparison shows that Coble creep in UN can be more dominant than (or is at least competitive with) dislocation creep. This agrees with the qualitative analysis by Konovalov \textit{et al.} \cite{Konovalov2016}. It should be mentioned that Galvin \textit{et al.} \cite{Galvin2025} used a Coble creep prefactor $A_C = 42 \pi$ for their UO$_2$ Coble creep formula, whereas in this work, we use $A_C = 46.347$. However, the difference between the two values is only a factor of 3, which would have little effect on the results in \cref{Fig:UNvsUO2}.

Merely comparing the properties of different fuel systems within a temperature range might not accurately represent how the fuel systems behave in reactor conditions. More appropriately, we should compare fuel behaviors at representative center-line temperatures \cite{Cheniour2020}. Considering the high thermal conductivity of UN, it is expected that the center-line temperatures and temperature gradients of UN fuel rods are much lower than those of UO$_2$ fuel rods under the same conditions. For example, based on their finite-element analysis of the thermal-mechanical behavior of several advanced fuel-cladding combinations, Zeng \textit{et al.} \cite{Zeng2021} have estimated the early-life centerline temperature of the UO$_2$-FeCrAl fuel system to be about 1300 K, whereas it is only about 900 K for the UN-FeCrAl fuel system. Thus, under the same early-life reactor conditions, the creep of UN is expected to be less pronounced than that of UO$_2$. The effect of irradiation on this comparison, however, requires a separate study.

\section{Discussion}


Parameters in \cref{Tab:CreepParams} calculated from the fitting of the diffusional creep simulation have about 16-24\% uncertainty, which is reasonable considering that each strain rate is an average over three unique simulations. The uncertainty can be reduced by simulating for longer times or averaging over more runs. However, this is computationally expensive for the supercells at hand, whose number of atoms ranges between 1,547,344 atoms for the 14-nm polycrystal and 4,537,943 atoms for the 20-nm polycrystal. Due to the much lower computational cost and smaller uncertainties, the parameter-based Coble creep model is recommended for use in the determination of a Coble creep rate.

While the simulations presented in this work utilize nanometer-scale polycrystals (14--20 nm) to achieve sufficient creep strain within accessible MD time scales, the fundamental deformation mechanism, namely, diffusional creep mediated by GB diffusion, is governed by well-established scaling laws. The Coble creep model inherently accounts for grain size effects, combined with the several hundred MPa stresses, through its $\sigma/d^{3}$ dependence, enabling extrapolation of MD-obtained parameters to realistic microstructural scales (on the order of micrometers) and typical creep stresses (tens of MPa).

An important aspect of this work is that we investigated the temperature dependence of the diffusional GB width and proposed a consistent method to calculate an effective GB width for the Coble creep formula. We found that the diffusional GB width in UN follows an Arrhenius behavior with an activation energy of about $Q_\delta$ = 0.74--0.82 eV. Keblinski \textit{et al.} \cite{Keblinski1999} assumed that the activation energy of GB diffusion ($Q$ = 2.28 eV in our case) can be expressed as $Q = Q_\delta + Q_\mathrm{amorph}$, where $Q_\mathrm{amorph}$ is the activation energy for diffusion in the amorphous structure of the GB. Based on this assumption, $Q_\mathrm{amorph}$ should be 1.47--1.55 eV. For this assumption to be verified experimentally, amorphous diffusion coefficients need to be measured for UN. However, it was never reported that UN amorphizes under irradiation. However, this does not affect our estimated GB width which nearly perfectly predicts the prefactor of the Coble creep formula, i.e., \cref{Eq:Coble}. Our two Coble creep models closely agree, especially at lower temperatures. However, considering that the parameter-based model generally exhibits smaller uncertainties in its calculated parameters, is much less computationally expensive, and can be more easily extended to a larger temperature range, we recommend its use in fuel performance codes. For ease of access, the complete Coble creep rate equation is provided here:

\begin{equation}
    \dot{\epsilon} = 582610.427 \, \frac{\sigma}{T d^3} \, \mathrm{exp} \! \left( - \frac{2.28 \text{ eV}}{k_\mathrm{B} T} \right),
    \label{Eq:OurModel}
\end{equation}
where $\sigma$ is the stress in MPa, $T$ is the temperature in K, $d$ is the average grain size in $\mu$m, and $\Dot{\epsilon}$ is the creep rate in s$^{-1}$. 

\begin{figure}[h!]
\centering
\begin{subfigure}{0.45\textwidth}
    \includegraphics[width=\textwidth]{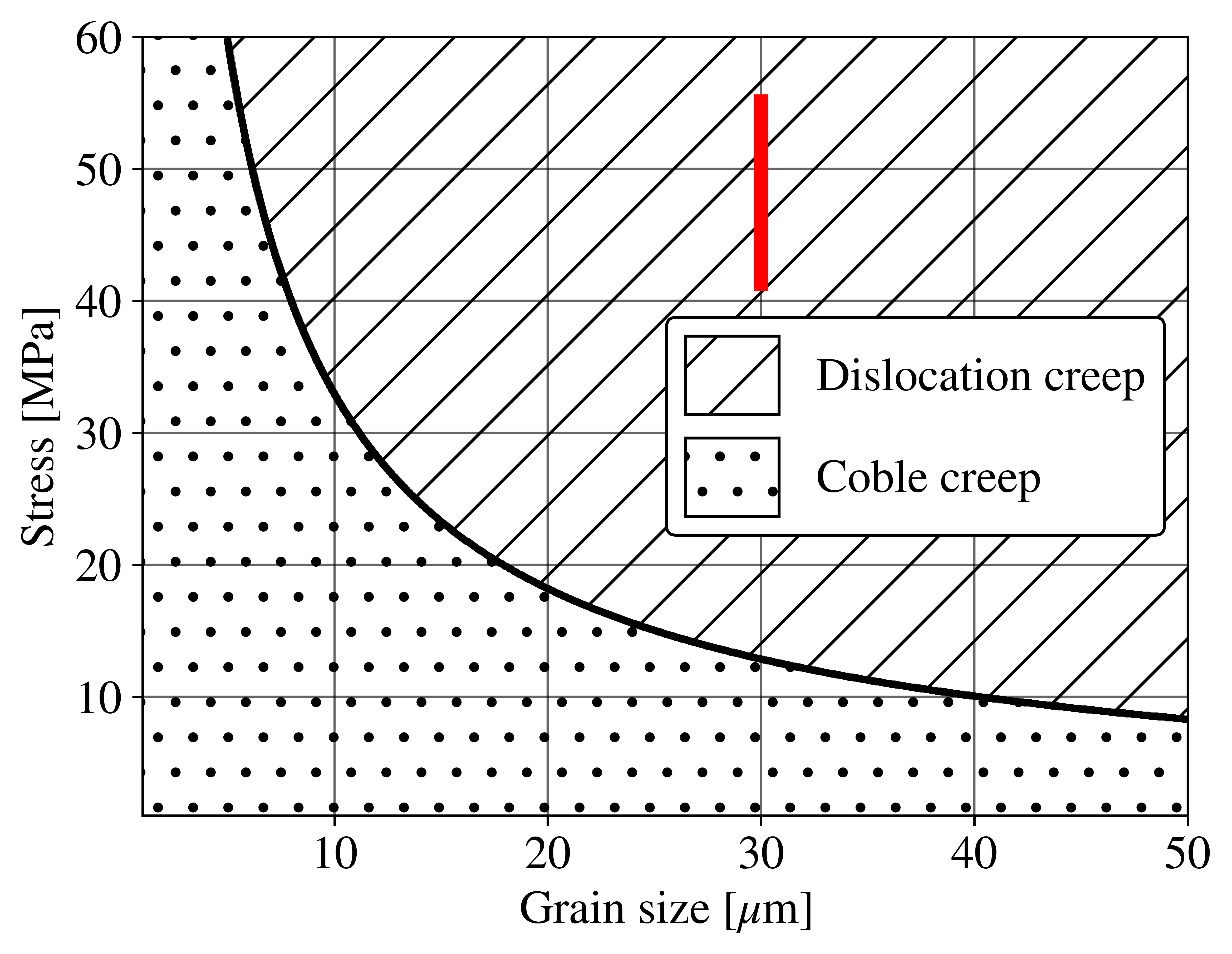}
    \caption{$T$ = 1800 K}
    \label{Fig:map1800}
\end{subfigure}
\hfill
\begin{subfigure}{0.45\textwidth}
    \includegraphics[width=\textwidth]{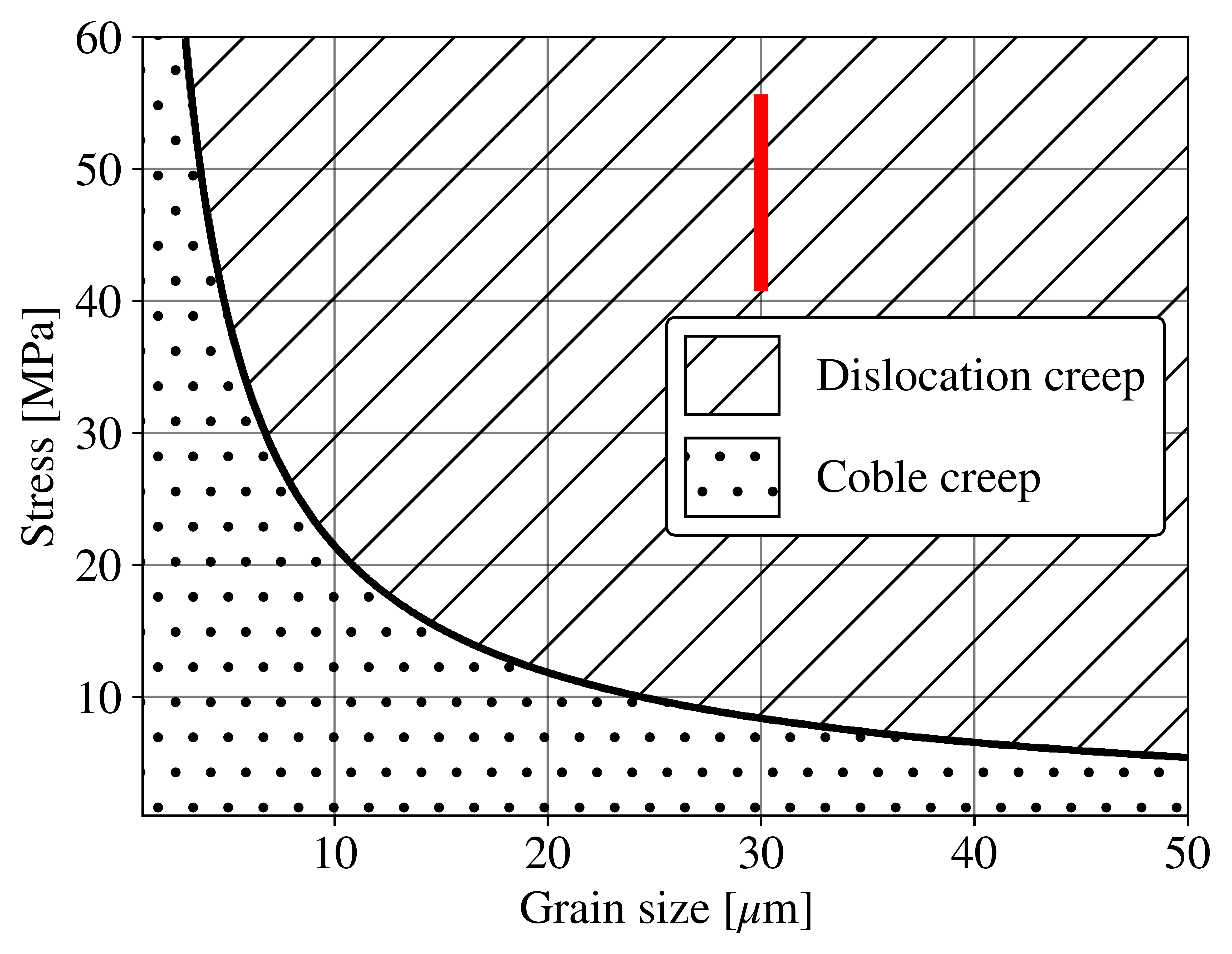}
    \caption{$T$ = 2200 K}
    \label{Fig:map2200}
\end{subfigure}
\hfill
\begin{subfigure}{0.45\textwidth}
    \includegraphics[width=\textwidth]{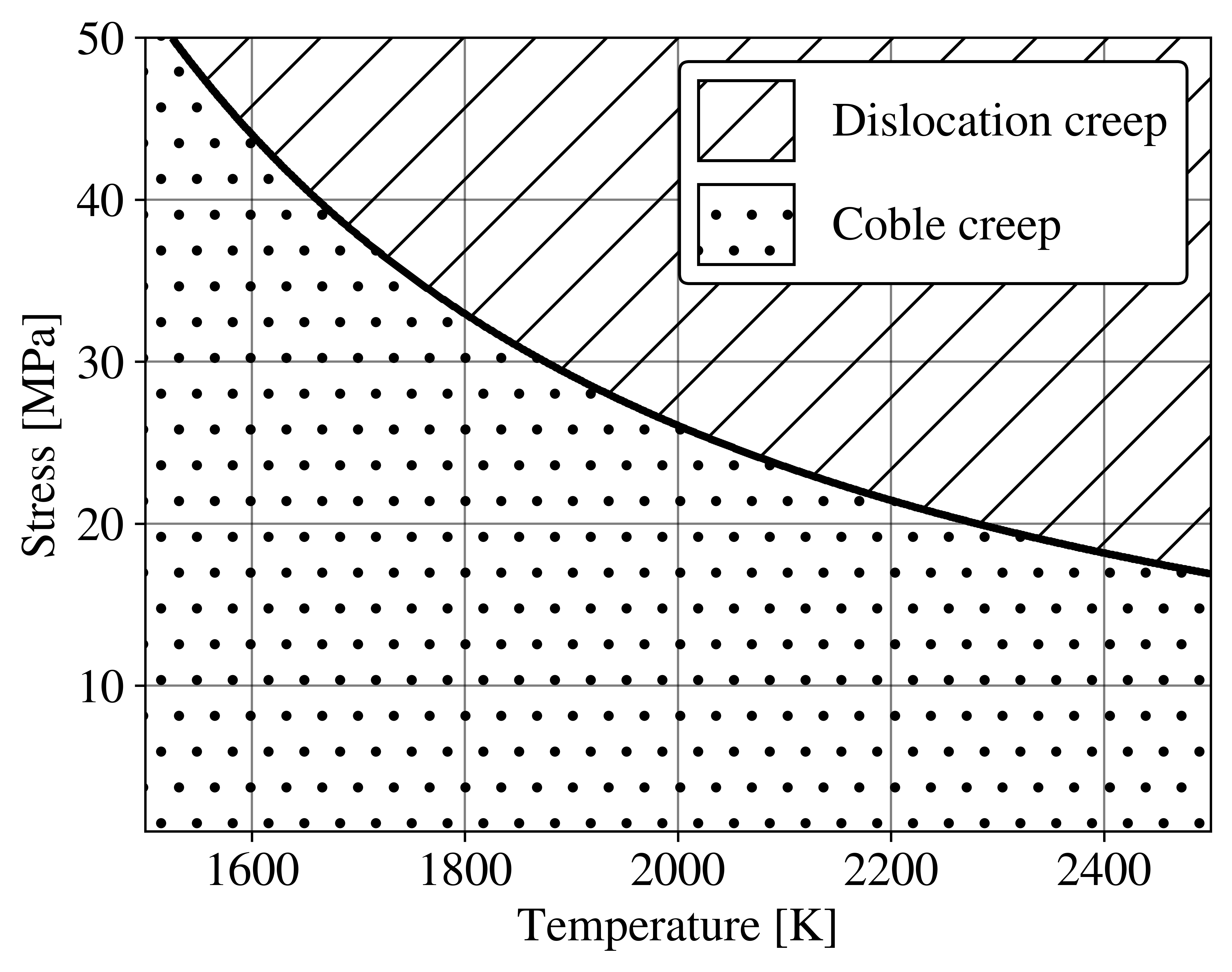}
    \caption{$d$ = 10 $\mu$m}
    \label{Fig:map10}
\end{subfigure}
\hfill
\begin{subfigure}{0.45\textwidth}
    \includegraphics[width=\textwidth]{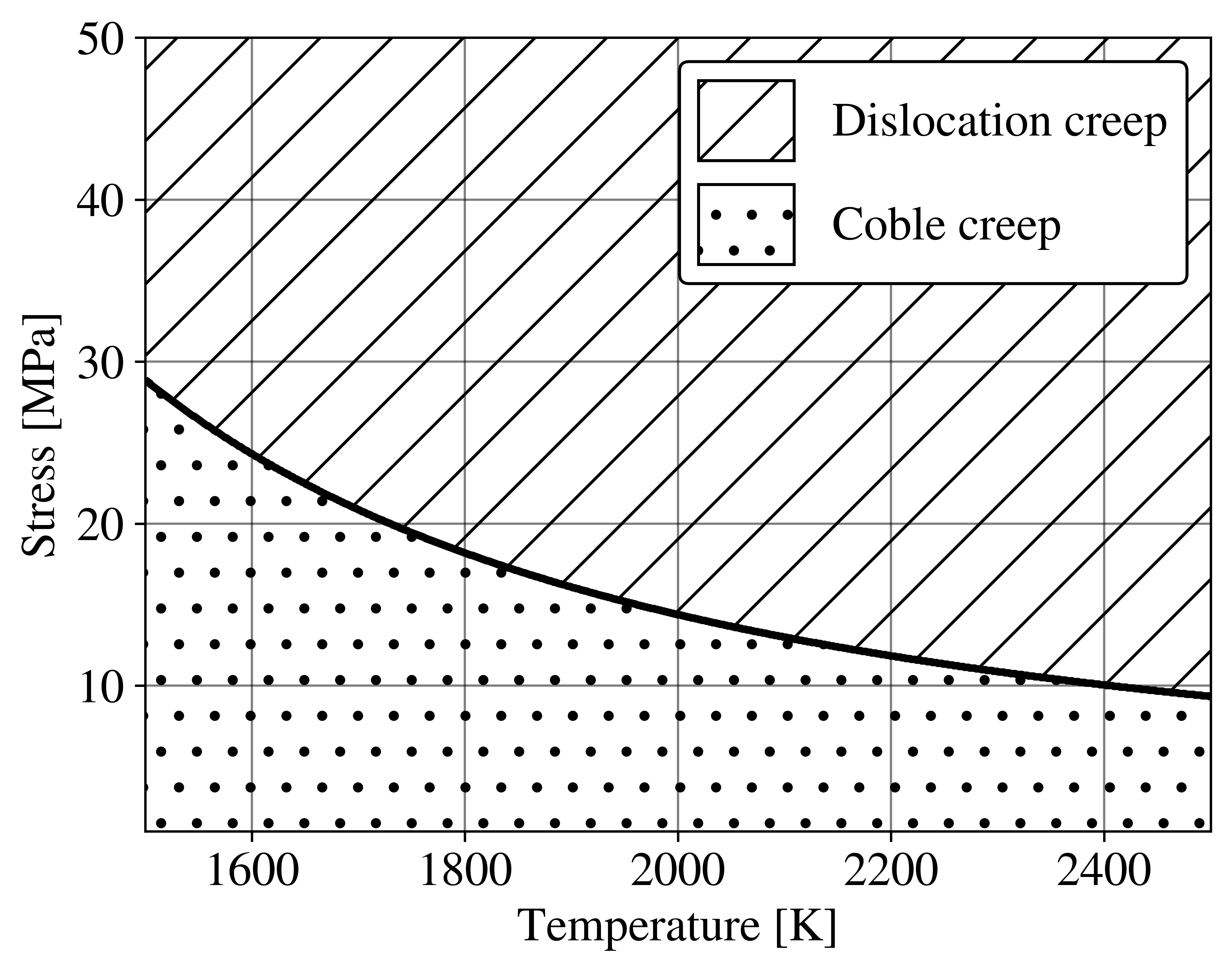}
    \caption{$d$ = 20 $\mu$m}
    \label{Fig:map20}
\end{subfigure}
\caption{(Color online) Deformation mechanism maps of the dominant creep mechanism for different grain sizes and stresses at \textbf{(a)} $T$ = 1800 K and \textbf{(b)} $T$ = 2200 K. The red solid line in \textbf{(a)} and \textbf{(b)} refers to the grain size and stress range of the creep rate measurements by Fassler \textit{et al.} \cite{Fassler1965}. Deformation mechanism maps of the dominant creep mechanism for different temperatures and stresses at \textbf{(c)} $d$ = 10 $\mu$m and \textbf{(d)} $d$ = 20 $\mu$m.}
\label{}
\end{figure}

The activation energy of Coble creep in UN is about 1 eV smaller than that of the dislocation-mediated creep estimated by Hayes \textit{et al.} \cite{Hayes1990II} and is very close to the activation energy of GB diffusion in UN assumed by Konovalov \textit{et al.} \cite{Konovalov2016}. This implies that there is a surface in the $\sigma, d, T$-space that separates regions in which Coble creep is dominant from other regions where dislocation creep is dominant. To examine this idea in some detail, we use \cref{Eq:Hayes,Eq:OurModel} (our Coble creep correlation and the Hayes dislocation creep correlation) to construct deformation mechanism maps for UN to identify regions where Coble creep is dominant and others where dislocation creep is dominant. Deformation mechanism maps that examine the dominant creep mechanisms for different grain sizes and stresses at $T$ = 1800 K and 2200 K are shown in \cref{Fig:map1800,Fig:map2200}, respectively. As expected, as the grain size is increased, dislocation creep becomes more dominant. The red solid line in \cref{Fig:map1800,Fig:map2200} refers to the grain size and stress range of the creep rate measurements by Fassler \textit{et al.} \cite{Fassler1965}, which had a stress exponent of about 4.7. The measurements are clearly in the dislocation creep region which explains the high stress exponent. It should be mentioned that Fassler \textit{et al.} data were obtained at 1373--1623 K.

\begin{figure}[h!]
    \centering
    \includegraphics[width=0.75\textwidth]{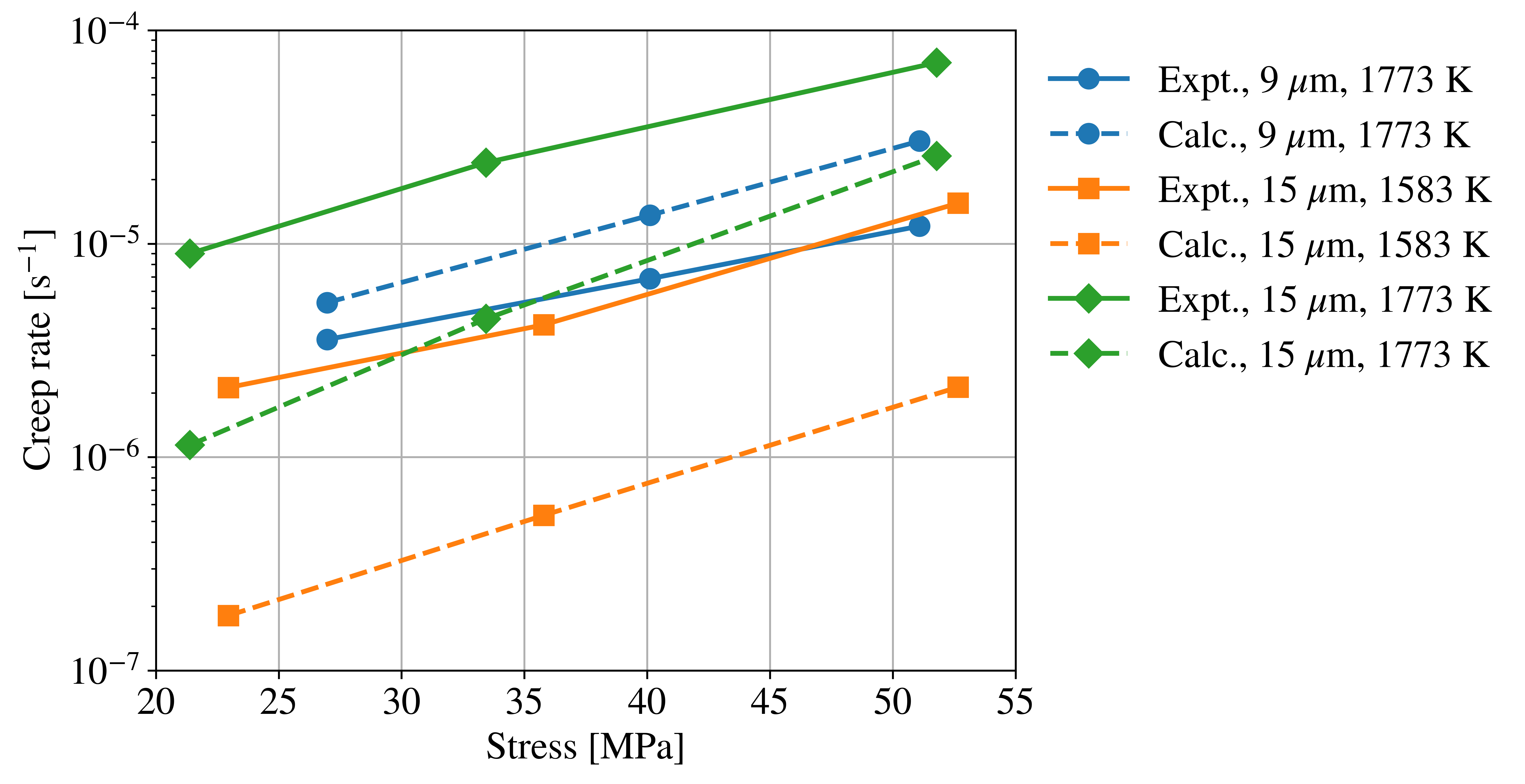}
    \caption{(Color online) Comparison of the experimental creep rate data of Uchida and Ichikawa \cite{Uchida1973} to that calculated using \cref{Eq:Final}.}
    \label{Fig:Final}
\end{figure}

Interestingly, at $d$ = 10 $\mu$m, the upper limit of the optimum UN grain sizes \cite{Johnson2018}, and the stress range of 20-30 MPa, within which the Hayes correlation was fitted, Coble creep is more dominant than dislocation creep at 1800 K, and both are nearly equally present at 2200 K. Thus, a contribution of both creep mechanisms will be observed with the contribution of the dislocation creep increasing with increasing temperature or increasing the grain size. Similar deformation mechanism maps for $d$ = 10 and 20 $\mu$m are shown in \cref{Fig:map10,Fig:map20}, respectively, for different temperatures and stresses. For a stress of 20 MPa and $d$ = 10 $\mu$m, dislocation creep becomes apparent at about $T$ = 2200 K, whereas for $d$ = 20 $\mu$m, Coble creep is only present at temperatures below $T$ = 1700 K, where the creep rate is small anyway. The most general thermal creep correlation for UN should be a sum of both contributions:
\begin{equation}
    \Dot{\epsilon}_\mathrm{tot} = 582610.427 \, \frac{\sigma}{T d^3} \, \mathrm{exp} \! \left( - \frac{2.28 \text{ eV}}{k_\mathrm{B} T} \right) + 2.054 \times 10^{-3} \sigma^{4.5} \, \mathrm{exp} \! \left( - \frac{3.39 \text{ eV}}{k_\mathrm{B} T} \right),
    \label{Eq:Final}
\end{equation}
where $T$ is in K, $\sigma$ is in MPa, and $d$ is in $\mu$m. To test this final correlation, we compare it to the experimental data of Uchida and Ichikawa \cite{Uchida1973} in \cref{Fig:Final}. For the grain size of 9 $\mu$m, the correlation gives an excellent prediction of the experimental creep rate. For the grain size of 15 $\mu$m, \cref{Eq:Final} underestimates the creep rate by at most an order of magnitude, which can be considered fair qualitative agreement taking into account that the 15-$\mu$m samples of Uchida and Ichikawa \cite{Uchida1973} experienced a significant hot-pressing effect.

While the present work addresses thermally activated diffusional creep in pristine UN, irradiation is expected to significantly influence creep behavior under reactor-relevant conditions. In particular, irradiation-enhanced diffusion can markedly increase atomic transport rates and enable creep rates far exceeding those predicted from thermal diffusion alone~\cite{Cooper2021}. Consequently, the results presented here should be interpreted as a lower bound on in-service creep behavior and motivate future studies that explicitly incorporate radiation damage.


\FloatBarrier

\section{Conclusions}

In this study, we employed MD simulations to investigate specific aspects of the mechanical behavior of UN. The diffusional creep simulations were conducted using nanometer-sized polycrystals with realistic microstructural features. Results indicate that Nabarro-Herring creep is not active within the studied temperature range and time scale. Instead, Coble creep emerges as the dominant diffusional creep mechanism, exhibiting an activation energy of 2.28 eV. A method is introduced to compute the diffusional GB width and its temperature dependence in polycrystals. The diffusional GB width of UN is observed to follow an Arrhenius behavior with an activation energy of 0.74--0.82 eV. The value of the diffusional GB width at the melting point is estimated as 2.69 nm, which we treat as an effective GB width to be used in the Coble creep formula. This value shows an excellent prediction of the prefactor of the phenomenological Coble creep formula. The parameter-based model of Coble creep in UN agrees with the simulation model and shows less uncertainty at less computational cost. It is shown that the most general thermal creep correlation of UN is a sum of our Coble creep correlation and the Hayes \textit{et al.} dislocation creep correlation.


\section{Acknowledgements}

The authors thank Khadija Mahbuba and Mahmoud Hawary for the fruitful discussions and useful suggestions. This work is funded by Westinghouse Electric Company. This research made use of the resources of the High-Performance Computing Center at Idaho National Laboratory, which is supported by the Office of Nuclear Energy of the U.S. Department of Energy and the Nuclear Science User Facilities under Contract No. DE-AC07-05ID14517. Mohamed AbdulHameed dedicates this work to Ayman AbdulRaheem.

\appendix

\section{Derivation of the diffusional GB width formula}
\label{app}

An estimate of the diffusional GB width can be made based on the formula \cite{Keblinski1999}:
\begin{equation}
\delta_d = \frac{N_{\mathrm{GB}} \Omega}{A} 
\label{Eq:dNGB}
\end{equation}
where $N_{\mathrm{GB}}$ is the number of atoms existing within the GBs, $\Omega$ is the atomic volume, and $A$ is the GB area. The number of GB atoms can be defined as $N_\mathrm{GB} = c N$ \cite{Haslam2004} where $c$ is defined as the fraction of atoms that have undergone non-affine displacements of at least one nearest-neighbor distance in the limit of infinite simulation time, and $N$ is the total number of atoms in the supercell. The atomic volume is $\Omega = gV/N$, where $g$ is the number of grains in the supercell (in our case, $g$ = 16), and $V$ is the grain volume. Because the grains are space-filling, the supercell volume $l_x l_y l_z = gV$. The GB area is $A = g S / 2$, where $S$ is the grain surface area, which for a truncated octahedron is given by \cite{Weisstein}:
\begin{equation}
S = \left( 6 + 12 \sqrt{3} \right) l^2    
\end{equation}
where $l$ is the truncated octahedron's edge length, which can be calculated from \cite{Weisstein}:
\begin{equation}
l = \left( \frac{V}{8 \sqrt{2}} \right)^{1/3}
\end{equation}
Thus, for the case of a supercell containing truncated-octahedral grains of uniform size and shape, the diffusional GB width is:
\begin{equation}
\delta_d = 2 c \frac{V}{S} = \eta c V^{1/3}
\label{Eq:VS}
\end{equation}
where $\eta$ is a shape factor that depends on the grain shape. For truncated-octahedral grains:
\begin{equation}\label{eq:shape}
\eta = \frac{4 \times 2^{1/3}}{3 + 6 \sqrt{3}} \approx 0.376.
\end{equation}
This derivation assumes that the grains are space-filling and of uniform shape and size. It is interesting to note in \cref{Eq:VS} that while the GB width, $\delta$, depends on the grain volume-to-surface ratio, $V/S$, which increases with increasing the grain size, $\delta$ also depends on $c$, which increases both with increasing the temperature and the grain surface-to-volume ratio. That is, this definition of the GB width takes into account both structural and diffusional aspects.

\bibliographystyle{elsarticle-num}
\bibliography{ref}

\end{document}